\documentclass[sigconf, nonacm]{acmart}

\AtBeginDocument{%
  }

\usepackage{multirow,multicol}
\usepackage{tabularx}
\usepackage[table]{xcolor}
\usepackage{subcaption}
\usepackage{soul}
\usepackage{tikz}
\usepackage[most]{tcolorbox}
\usepackage{enumitem}
\usepackage{wrapfig}

\newcommand{\circled}[1]{\tikz[baseline=(char.base)]{
            \node[shape=circle,fill,inner sep=0pt] (char) {\textcolor{white}{#1}};}}

\begin{document}

\title{Distributed Persistence Domain for Persistent Memory Pooling}

\author{Khan Shaikhul Hadi}
\email{shaikhulhadi@ucf.edu}
\affiliation{%
  \institution{University of Central Florida}
  \city{Orlando}
  \country{FL, USA}
}

\author{Andres David Delgado}
\email{andres.delgado@ucf.edu}
\affiliation{%
  \institution{University of Central Florida}
  \city{Orlando}
  \country{FL, USA}
}
\author{Naveed Ul Mustafa }
\email{num@nmsu.edu}
\affiliation{%
  \institution{New Mexico State University }
  \city{Las Cruces}
  \country{NM, USA}
}
\author{Mark Heinrich}
\email{heinrich@ucf.edu}
\affiliation{%
  \institution{University of Central Florida}
  \city{Orlando}
  \country{FL, USA}
}
\author{Hao Zheng }
\email{hao.zheng@ucf.edu}
\affiliation{%
  \institution{University of Central Florida}
  \city{Orlando}
  \country{FL, USA}
}
\author{Yan Solihin}
\email{yan.solihin@ucf.edu}
\affiliation{%
  \institution{University of Central Florida}
  \city{Orlando}
  \country{FL, USA}
}


\begin{abstract}

Compute Express Link (CXL) enables memory pooling over disaggregated memory, offering the potential to improve resource utilization in persistent memory systems. However, integrating persistence semantics into CXL-based memory pooling introduces substantial latency, which limits system scalability. This overhead arises because persist operations must traverse the entire CXL fabric, including switches, links, and protocol layers, before reaching remote persistent memory. To this end, we argue that extending CXL switches with persistence support is a promising direction for improving the scalability of persistent memory pooling. However, moving persistence support into the network breaks the traditional correctness assumptions of centralized persistence domains. In particular, enabling persistence within distributed structures, such as CXL switches, can introduce stale reads and writes if not carefully coordinated.

In this paper, we propose Distributed Persistence Domain (DPD), a new abstraction for persistent memory pooling that enables persistence support at the CXL switch level. We first formalize the concept of a distributed persistence domain and use DPD as a framework to identify the correctness hazards that arise when persistence structures are distributed across the CXL fabric. Based on this analysis, we derive the design requirements needed to guarantee correctness. Building on these insights, we present Persistent CXL Switch, a CXL switch architecture that incorporates persistence support to significantly reduce persist latency, enable read forwarding, and coalesce writes, while preserving correctness and crash consistency. We evaluated our system design using both SPLASH-4 and YCSB benchmarks. Simulation results show an average speedup of 33\% over volatile CXL switches, and up to 36\% speedup with read forwarding optimization across all workloads.

\end{abstract}


\maketitle

\section{Introduction}

Compute Express Link (CXL) \cite{sharma2024cxl_introduction,sullivan2022cxl_fabric_introduction} has emerged as a promising technique for memory expansion\cite{li2025performance_cxl, sun2023demystify_cxl} and memory disaggregation \cite{boles2023cxl_enhanced_function,sharma2022datacentric_compute} with memory pooling\cite{zhang2020rpma,pond_2023} and multi-host memory sharing  ~\cite{huang2025tigon, ma2024hydraRPC, yang2025polarCXLmem, yang2024polarDPMP} capability. By attaching memory devices through CXL protocols, systems can decouple memory capacity from compute nodes and enable high utilization and flexible provisioning for HPC \cite{fridman2023pm_disaggregation_hpc, ding2024evaluating_dm_hpc, fridman2021use_pm_hpc} and data-intensive workloads \cite{ruan2023pm_disaggregation_for_cloud, sim2023CMS}. Given their substantial advantage in capacity and density, persistent memory devices will likely be attached to CXL, e.g. 1-4TB Samsung CMM-H~\cite{zeng2025samsung_cmmh}, creating a persistent disaggregated memory system to support memory pooling. However, when persistent memory becomes physically distributed across the CXL fabric\cite{sullivan2021cxl_support_pm, ruan2023pm_disaggregation_for_cloud, tsai2020disaggregating}, the system must preserve crash consistency and data correctness across multiple intermediate, rather than a single, co-located persistence  domain. This raises new challenges in sustaining performance scalability while ensuring correctness.

\begin{figure}[!t]
    \centering
    \includegraphics[width=\linewidth]{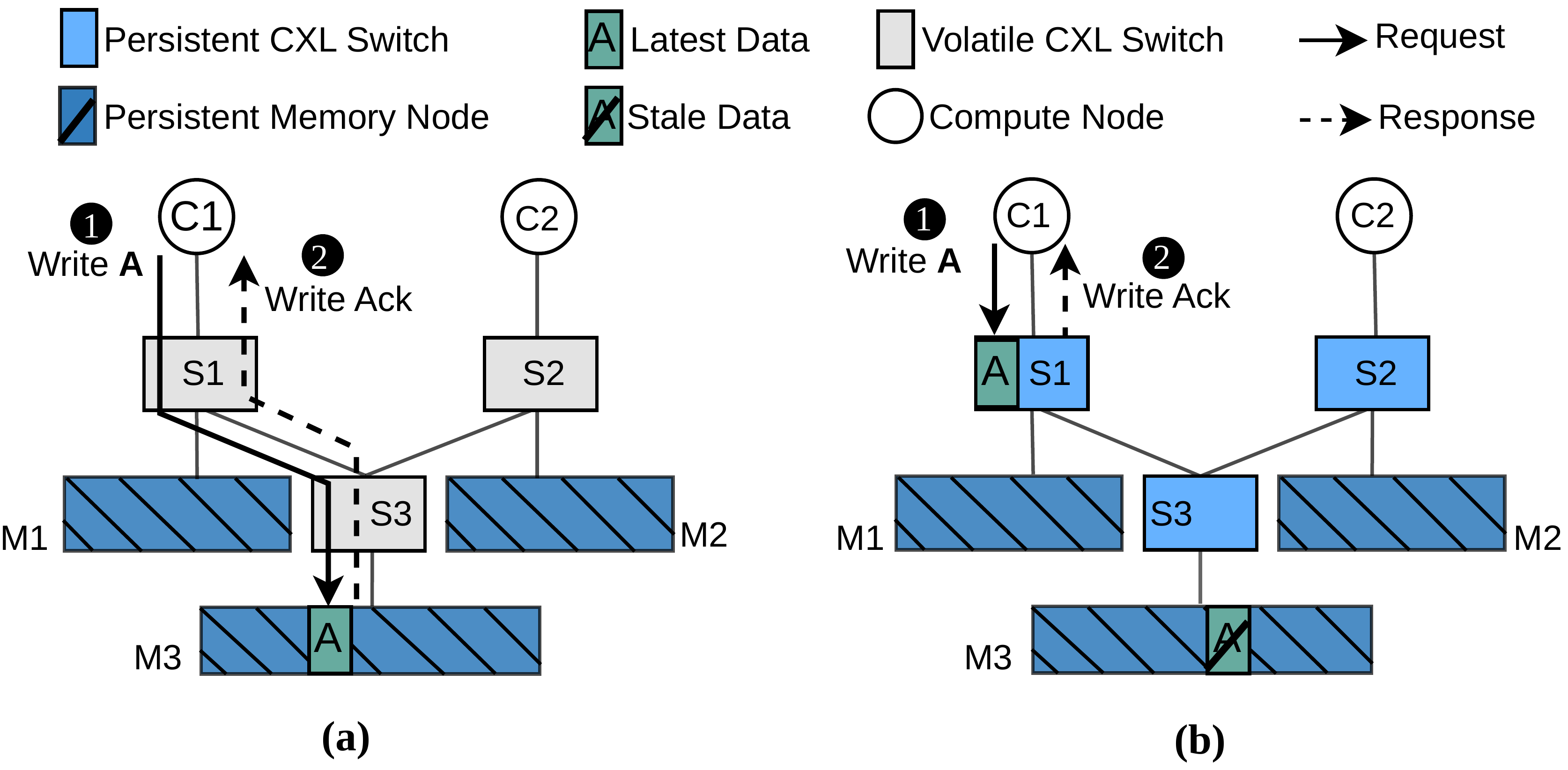}
    \vspace{-2em}
    \caption{ (a) Persistent memory pooling using volatile CXL switch, and (b) Proposed persistent CXL switch.}
    \label{fig:cxl_persistence_diagram}
    \vspace{-1em}
\end{figure}

While substantial efforts~\cite{eADR,alshboul-hpca21-bbb, zeng2023persitent_process, narayanan2012wsp} have extended the persistence domain within the memory hierarchy, they are optimized to systems with a single processor-attached persistence domain. Naively adopting these designs for pooling in a disaggregated persistent memory setting  would result in unacceptable latency overheads. For example, BBB\cite{alshboul-hpca21-bbb} closes the gap between cache and persistent memory by providing a battery-backed persist buffer, enabling strict persistency and eliminating costly flush/fence instructions. Horus~\cite{han2022horus} tackles secure extended persistence domains by optimizing draining of metadata and integrity-tree updates inside eADR-like systems. Likewise, non-volatile caches (NV-Caches) \cite{izraelevitz2016failure, youyou_iccd2014_loose_ordering} and various crash-consistency frameworks~\cite{coburn2011nv_heap, joshi2015eff_persist_barrier, joshi2017atom, hadi2023DAI, sihang_asplos2023_cross_failor} focus on ensuring correctness when persistence is local to a CPU and its immediate memory controller. Prior efforts all rely on the assumption that the persistence domain forms a continuous path from the point of persistence initiation to the final persistent memory module. Applying this correctness assumption to a disaggregated persistent memory system mandates enforcing persistence at the remote endpoint, where persist latency increases proportionally with system scale as shown in Fig. ~\ref{fig:cxl_persistence_diagram}(a).

Decoupling persist latency from system scale requires moving persistence capabilities into the network layer \cite{seemakhupt_isca2021_pmnet, hu2018persistence_parallelism}, such as the CXL switch. However, maintaining correctness in a persistent CXL switch is challenging because the ordering of persist writes and reads cannot be guaranteed across the network fabric \cite{karim_tos2025_nvm, assa_arx2025_programming_cmd}. Prior work on disaggregated persistent memory primarily focuses on the distributed nature of memory nodes and uses software techniques to achieve crash consistency and improve performance. Hotpot~\cite{shan2017hotpot} develops a custom kernel for distributed shared persistent memory, providing low-latency and transparent memory access, along with data persistence and reliability. Ethane~\cite{cai2024ethane} improves performance by optimizing memory access through an asymmetric file system, while AsymNVM~\cite{ma2020asymnvm} redesigns the memory architecture asymmetrically to further enhance efficiency. FORD~\cite{zhang2022ford} reduces the number of requests required to ensure crash consistency across distributed persistent memory nodes, thereby improving performance. Dinomo~\cite{lee2022dinomo}, RDMP-KV~\cite{li_sc2020_rdmp}, and pDPM~\cite{tsai2020disaggregating} introduce distributed persistent-memory-aware data structures to accelerate operations. To the best of our knowledge, no prior work has attempted to extend the persistence domain in disaggregated persistent memory systems in a distributed manner to improve the scalability of persistent operations while simultaneously guaranteeing correctness and crash consistency for persistent memory pooling.

Enabling persistence support within the network layer can create multiple versions of the same data across the CXL fabric. A conventional solution is to employ coherence-like mechanisms that actively invalidate stale data copies~\cite{sorin2011primer, yeap2025riscv_cache}, or to rely on self-invalidation techniques to enforce consistency~\cite{singh2013gpu_coherence, ashby2011bloom_filter, kaxiras2012snoopless_coherence, kaxiras2015coherence_DSM, lebeck1995dynamic_self_invalidation}. However, naively adopting existing coherence protocols can undermine persist performance~\cite{lebeck1995dynamic_self_invalidation, kaxiras2010micro_SARC, hechtman2012concurrency_limits}. This calls for a careful examination of correctness mechanisms that preserve data consistency without sacrificing the performance benefits of adding persistence to CXL switches.

In this paper, we introduce Distributed Persistence Domain (DPD), a new abstraction that moves persistence support into the network layer to reduce persist latency, as shown in Fig.~\ref{fig:cxl_persistence_diagram}(b). DPD can be used to allow multiple copies and versions of data to exist across distributed persistence structures, such as CXL switches and memory controllers, while preserving correctness. DPD does not require the entire network to be persistent; any subset of switches can be part of the persistence domain, regardless of their location in the fabric. DPD is orthogonal to prior processor-centric and software-based techniques, and can be combined with them to further enhance persistent memory performance.

The major contributions of this paper are:

\begin{itemize}[topsep=1.5em]
    \item \textbf{Distributed Persistence Domain.} We propose \textit{Distributed Persistence Domain} (DPD), a new approach to define persistence domain for disaggregated persistent memory that enables persistence at the network fabric rather than being enforced at remote memory. The proposed concept provides a foundation to add data persist capability for sustaining performance scalability in persistent memory pooling.

    \item \textbf{Persistent CXL Switch Design.} We propose a CXL switch design with built-in data persistence capability to demonstrate the feasibility of our approach. DPD is used as a framework to determine design criteria for the proposed persistent CXL switch (PCS). The proposed PCS can reduce persist latency, enable read forwarding, and coalesce writes to alleviate bandwidth demand, while ensuring data correctness and crash consistency for pooling across disaggregated persistent memory systems.
    
     \item \textbf{Correctness Guarantee for Distributed Persistence Domain:} We examine the correctness challenges when adopting a distributed persistence domain, and show that with minimal software support, PCS can ensure both data correctness and crash consistency.
\end{itemize}

We evaluate DPD using the Splash-4~\cite{splash4} parallel benchmark suite and YCSB \cite{cooper2010ycsb} with memcached \cite{fitzpatrick2011memcached} database adapted to persistent memory. While preserving correctness and crash consistency, our results show that ensuring persistence at PCS alone delivers a 33\% performance improvement over a baseline system without PCS support. When enabling read forwarding at PCS, the average speedup further increases to 36\% .



\section{Background and Motivation}

\subsection{Compute Express Link}
Compute Express Link (CXL) \cite{sharma2022cxl_openstandard, sharma2024cxl_introduction} is an open industry-standard interconnect protocol  developed to facilitate high-bandwidth, low-latency communication between processors (called hosts) and devices such as memory devices and accelerators. CXL enables memory expansion, and memory disaggregation with memory pooling~\cite{li2025performance_cxl, sun2023demystify_cxl, sharma2022datacentric_compute, sharma2024cxl_introduction, boles2023cxl_enhanced_function, zhang2020rpma,pond_2023}. The CXL protocol consists of CXL.io, CXL.cache and CXL.mem which are dynamically multiplexed over the PCIe physical layer. CXL.cache is used to maintain coherence in shared data, CXL.mem is used to read and write 64-byte cache line data from memory and CXL.io is based on PCIe mechanisms for I/O communication\cite{boles2023cxl_enhanced_function}. CXL devices can be one of three types: Type 1 devices can access and cache host memory, Type 2 devices can coherently share their own memory with the host, and Type 3 devices are only load/store accessible (i.e. CXL-attached memory). Since CXL v1.0, coherence and memory semantics have been supported~\cite{sharma2022cxl_openstandard, rudoff2021cxl_pm} for devices including persistent memory (e.g. memory-semantic SSD~\cite{samsung_memory_semantinc_ssd} and Optane Pmem~\cite{optane_brief}). This work focuses on CXL.mem with attached PM devices. 

\vspace{-0.5em}
\subsection{Persistent Memory  Disaggregation}
Persistent memory pooling allows a host to expand its memory capacity on demand from an available disaggregated persistent memory pool~\cite{lee2023PACT_sdm,tsai2020disaggregating}. Current memory disaggregation practices rely on RDMA over Ethernet and InfiniBand, where Storage Area Networks (SANs) provide remote memory access with latencies of 1--2 microseconds~\cite{juncheng2017infiniswap, anju2014rdma_keyvalue}. However, RDMA-based pooling faces several critical limitations for persistent memory: (1) microsecond-scale latency for latency-sensitive persist operations~\cite{tsai2020disaggregating}, (2) variable performance in multi-tenant environments due to network contention~\cite{groves2017unraveling_network_memory_contention}, and (3) explicit data movement operations that conflict with persistent memory programming models, which rely on cache-line flush and fence instructions rather than RDMA verbs~\cite{tsai2020disaggregating,duan2021hardware_rdma_pm}. CXL addresses these limitations by providing cache-coherent, load/store-accessible memory pooling with significantly lower latency (170--400 ns)~\cite{maruf2023memory_disaggregation}. The CXL protocol was also designed to support persistent memory (PM) devices, including features for persistency management, such as global persistent flush~\cite{sullivan2021cxl_support_pm, fridman2023pm_disaggregation_hpc}. In addition, CXL 3.0 introduces switched-fabric topologies that enable scalable memory pooling within and across racks~\cite{sullivan2022cxl_fabric_introduction}.

However, disaggregating persistent memory introduces new correctness and performance challenges, including compromised durability of in-flight writes, ambiguity in persist ordering due to routing, and the need to coordinate flushes across compute nodes~\cite{ patil-dsn23-apta, ruan2023pm_disaggregation_for_cloud}. To address these issues, Apta~\cite{patil-dsn23-apta} proposes a fault-tolerant coherence protocol to ensure durability, while PilotDB~\cite{ruan2023pm_disaggregation_for_cloud} targets relational databases and introduces compute-side logging and log-pulling techniques to reduce remote persist latency and improve recovery speed. However, both works either focus on higher-level software mechanisms or assume limited complexity in the persist path in shared memory.

\subsection{Persistence Domain}

The persistence domain (PD) refers to the set of hardware components that can accept a persist operation and guarantee data durability, ensuring that data survives power failures and system crashes~\cite{rudoff2017persistent_programming}. The boundary of PD evolves over time, as conventional systems with locally-attached PM define various boundaries of the PD. Intel's Asynchronous DRAM Refresh (ADR) requires the inclusion of memory controller's write pending queues in the PD\cite{PM_Architecture_book} while Enhanced ADR (eADR)\cite{eADR,rudoff2020persistent} expands the PD to include CPU caches. BBB ~\cite{alshboul-hpca21-bbb} adds a near-core persist buffer  to reduce required backup power while PPA~\cite{zeng2023persitent_process} aims to transparently extend the PD to cover all volatile processor state. All these works assume tightly integrated memory hierarchies with predictable latencies and a single coherence domain.

\subsection{Persistent Programming Model}
The use of persistent memory allows applications to manage crash consistency,  a property whereby, upon a crash (including power loss), the memory state remains consistent such that it allows the computation to recover from the crash. Crash consistency is maintained by explicitly specifying {\em persists}~\cite{rudoff2017persistent_programming}, which are stores that must reach durability before other memory accesses can proceed. Persists are supported differently across instruction sets. For example, Intel x86 provides \textit{clflush, clflushopt \& clwb} instructions to perform cache line flush or write-back and \textit{mfence, sfence} to ensure flushes reach the persistent domain at that point~\cite{intel64manual}. ARM ISA provides a combination of \textit{DC CVAP, DC CVADP} for write-back and \textit{DSB} as a fence~\cite{arm_manual}. Those instructions may be used in a library (e.g., Intel PMDK\cite{pmemio}) or directly in the application program to enforce persists and achieve crash consistency. Fundamentally, a persist makes the program wait until data is written to PM before continuing, therefore persist latency is critical to performance.



\subsection{Motivation}


\label{label:motivation}

\begin{figure}[!t]
    \centering
    \includegraphics[width=\linewidth]{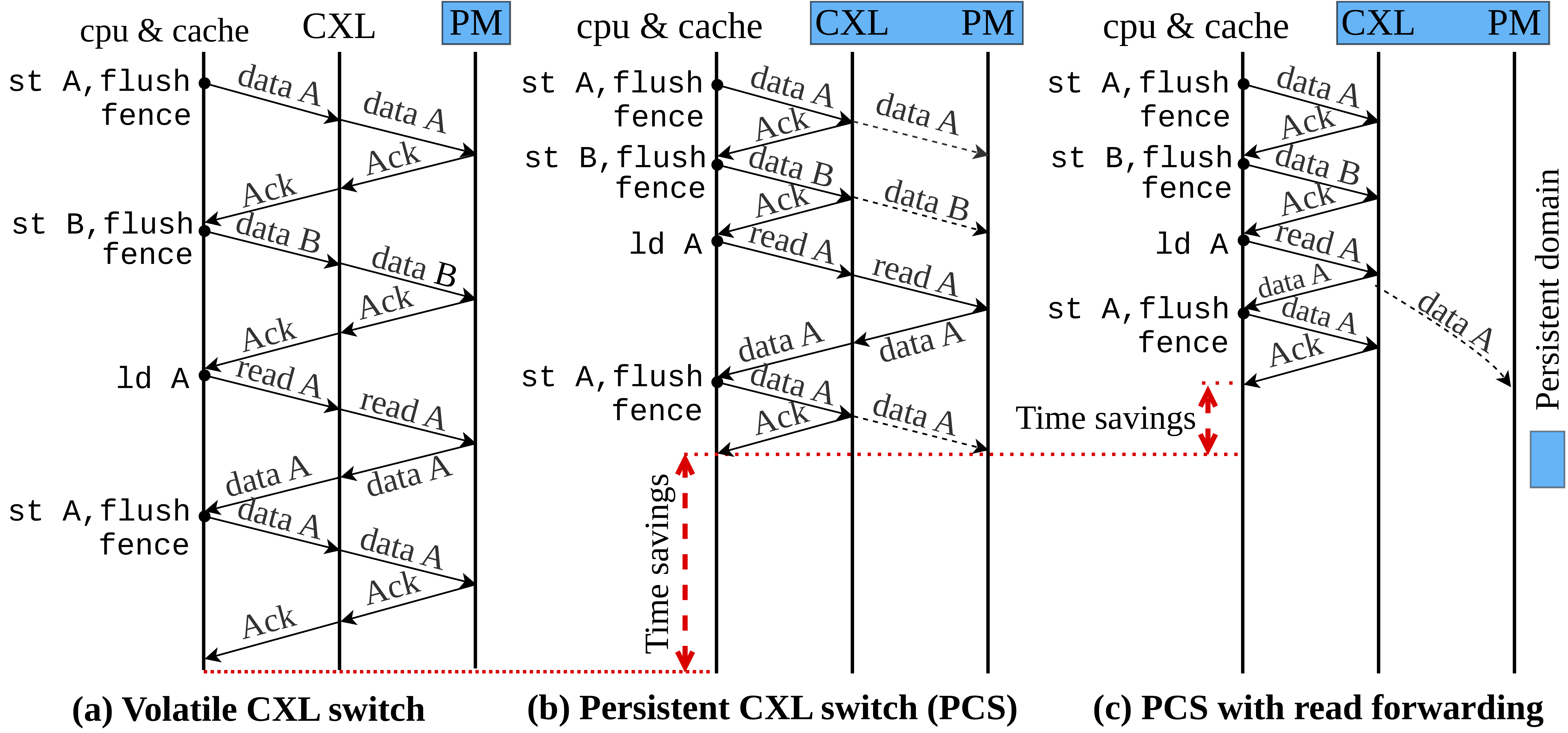}
    \vspace{-1.5em}
    \caption{(a) Persistent write and read in volatile CXL persistent memory pooling, (b) persist write operation at persistent CXL switch (PCS), (c) PCS with read forwarding.}
    \vspace{-1em}
    \label{fig:pcs_motivation}
    \vspace{-0.3em}
\end{figure}

As memory architectures evolve toward disaggregation to support memory pooling, we argue that PD needs to evolve toward disaggregation as well, where we define CXL-side PD that includes the CXL switch(es) and CXL-attached PMs. This is appropriate for several reasons. First, the memory access latencies depend heavily on the type of memory attached to the CXL and on the number of CXL switches involved. Thus, a CXL memory access latency may vary a lot, making it challenging for a processor-centric approach to provision an appropriate buffer size or energy capacity for ensuring persistency that works with a large memory latency variation. Furthermore, processor architecture and CXL memories are unlikely to be co-designed hence and therefore have different design cycles. On the other hand, a CXL switch will be designed with the types of memories it will be attached to. Finally, a CXL memory pool will likely have a separate power supply than the CPU, thus creating separate power failure domains. A power failure at the CXL side may create complications for persisting processor-side stores.

To achieve a CXL-side PD, we must define semantics and hardware primitives to support distributed, multi-path, and failure-resilient persistency. In the context of memory pooling, persist operations must traverse the CXL interconnect fabric, exposing full round-trip latency to remote memory nodes as demonstrated in Fig. ~\ref{fig:cxl_persistence_diagram}(a). This significantly amplifies the performance impact compared to locally attached PM. For example, a persist that completes in 200-300ns locally could add ~200ns latency for each  switch  when traversing multiple CXL switches ~\cite{liu2025cxl_characterization}. Furthermore, with  a compute node accessing persistent data structures across multiple memory nodes within the memory pool through different CXL paths, the persistent programming model must account for varying persist latencies and potential reordering across different persistent structure within the persistence domain. For example, as shown in Fig.~\ref{fig:pcs_motivation}(a), the persist latency in a traditional CXL system is high. Each persist requires sending data to the CXL switch and PM, and waiting for an acknowledgment to return to the CPU before the fence can be retired from the processor pipeline. Persist operations may execute sequentially to guarantee crash consistency. Consequently, four full turnaround trips occur in this example.

\begin{wrapfigure}{r}{0.48\columnwidth}
    \centering
   \vspace{-1em}
    \includegraphics[width=\linewidth]{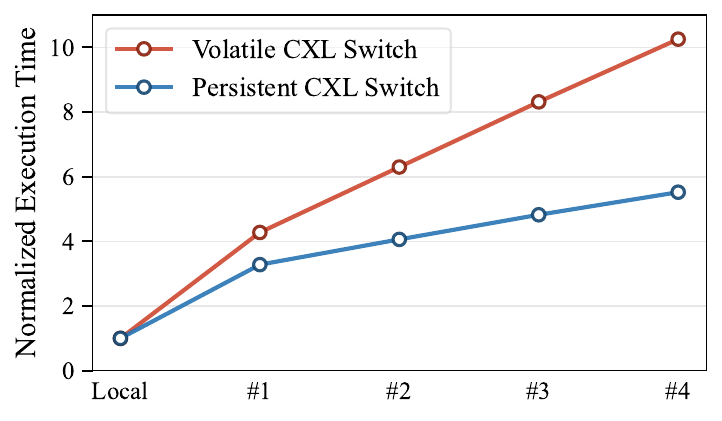}
   \vspace{-2em}
    \caption{Normalized execution time of FFT in current CXL persistent memory pooling, and with PCS across different hop counts. }
    \vspace{-1.2em}
    \label{fig:pcs_potential}
\end{wrapfigure}

To address this challenge, we argue that incorporating persistency support at the network level is crucial for future persistent memory pooling. As illustrated in Fig. \ref{fig:pcs_motivation}(b), enabling persist capability within the CXL switch can significantly reduce persist latency by allowing the switch to send an early acknowledgment to the CPU, rather than waiting for confirmation from the remote memory controller. By overlapping this acknowledgment with the subsequent transfer of data to the persistent memory, the critical path of the persist operations is shortened with considerable latency reductions. Furthermore, retaining data within the persistent switch enables read forwarding, eliminating the need to traverse the fabric and fetch the data copy from remote persistent memory. Based on our initial study using FFT applications, moving persistency from the memory endpoint to the network can achieve up to a 100\%  speedup, as shown in Fig.~\ref{fig:pcs_potential}. This holds significant promise for scaling performance in large-scale memory pooling, despite the new correctness challenges that we project.

 \begin{figure}[!t]
    \centering
    \includegraphics[width=\linewidth]{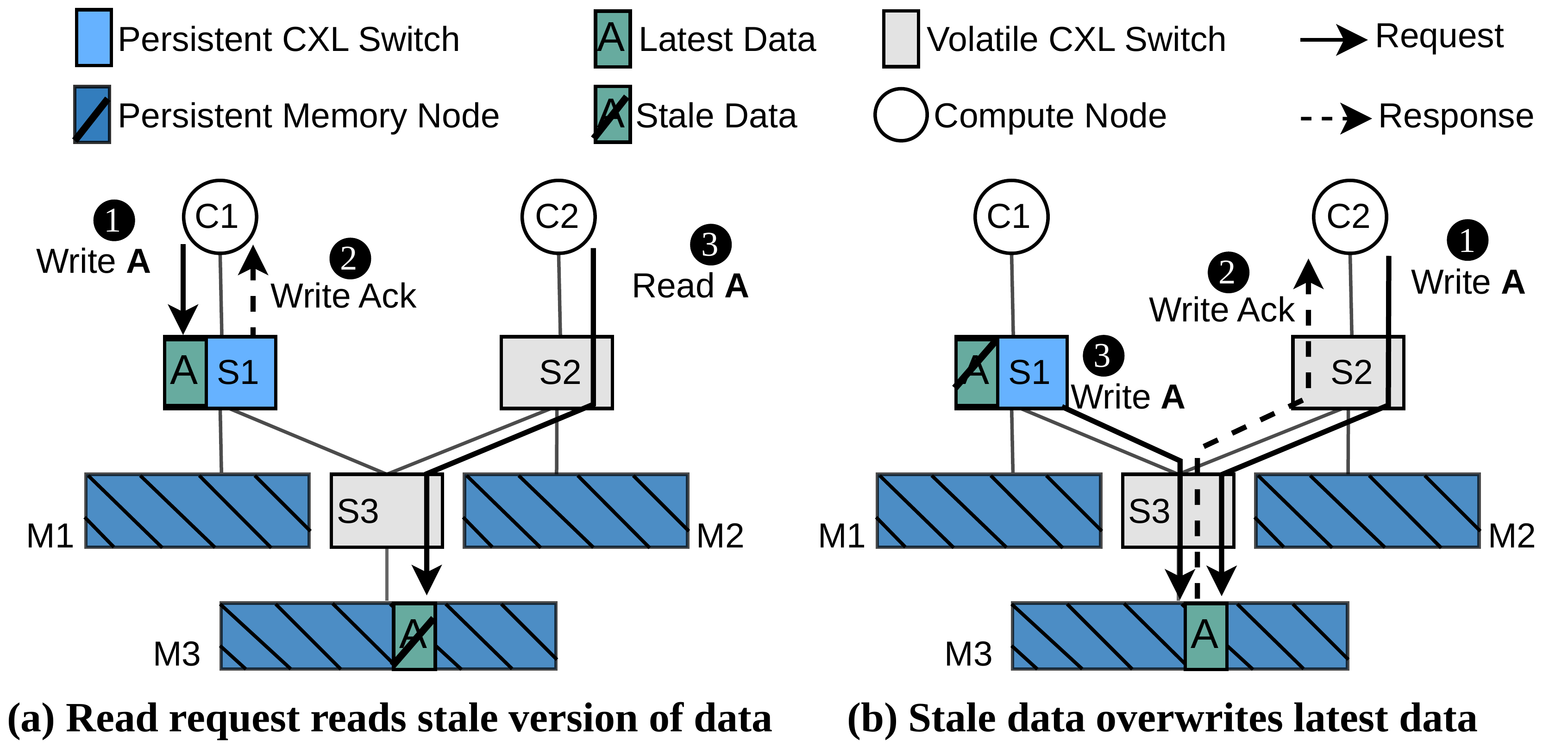}
    \vspace{-2em}
    \caption{Data correctness concern for persistent memory pooling with persistent CXL switch }
    \vspace{-1em}
    \label{fig:motivation_dpd}
\end{figure}




\vspace{-1em}
\section{Distributed Persistence Domain} 
A distributed persistence domain consists of multiple persistence structures, such as CXL switches and memory controllers, which together form an address-space persistence domain with multiple access points.


 \begin{figure*}[!t]
    \centering
    \includegraphics[width=0.95\textwidth]{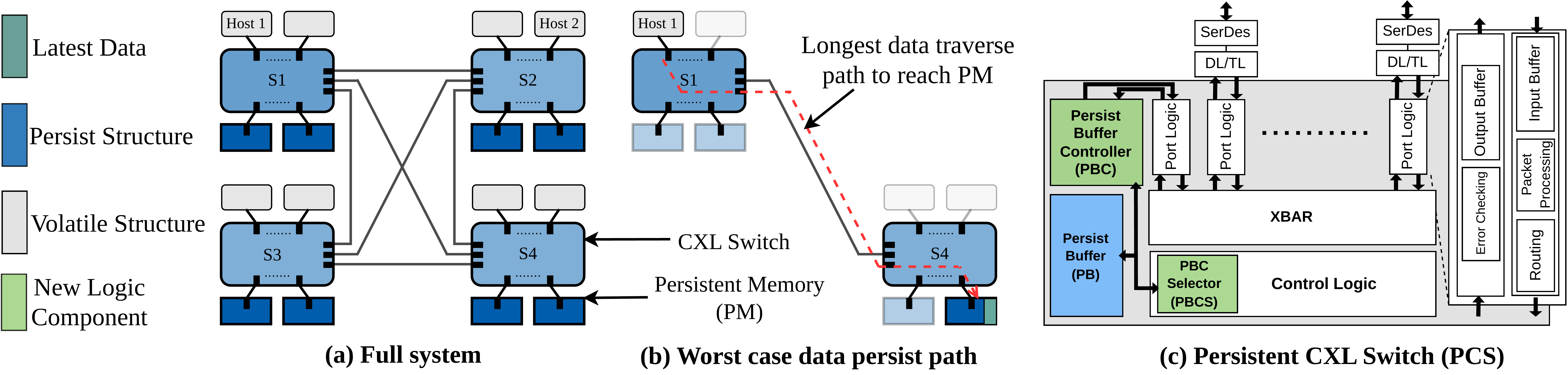}
    \vspace{-1em}
    \caption{ CXL-based disaggregated persistent memory.  System has multiple persist structures and multiple hosts that inject data (a).  For a specific host, data correctness depends only on the persist structures along the path data takes to reach persistent memory (b). CXL switch with persistent buffer allow data to be persisted at switch level (c).}
    \vspace{-1em}
    \label{fig:system_design}
\end{figure*}

\subsection{Correctness Requirement}

In the distributed persistence domain,  multiple versions of the same data could reside in different persistence structures, leading to stale-data reads and overwrite hazards. For example, as illustrated in Fig.~\ref{fig:motivation_dpd}(a), S1 may persist the most recent copy of the data, while a subsequent read request issued from a different requesting node (\circled{3}) may traverse a different path and ultimately read a stale copy residing in M3. Similarly, as shown in Fig.~\ref{fig:motivation_dpd}(b), the copy of the data at S1 may itself become stale after a new write is issued (\circled{1}) and eventually persisted in persistence structure M3. If S1 is unaware that its local copy is stale, it may later write that copy back to M3 (\circled{3}), leading to a write-reordering hazard. Furthermore, after a crash, the recovery process may be unable to determine the latest version of the data when multiple versions are present within the network.

Conventional systems typically rely on coherence protocols to enforce write propagation and serialization. However, such approaches are not well suited for ensuring correctness at the network level for several reasons. First, invalidation-based protocols require stale copies to be invalidated before a persist operation can complete, causing the persist operation to stall until invalidation finishes~\cite{huh2004coherence_decoupling, pan2014_coherence_miss_modeling}. This can significantly increase persist latency. Second, software-based synchronization and self-invalidation approaches require substantial modifications to applications and processors in order to forward coherence requests from the processor side to the network side~\cite{choi2011_denovo, tavarageri2016_coherence_compiler}. Third, these mechanisms introduce additional network traffic, such as invalidation messages, which reduces the bandwidth available for data transfer~\cite{gupta1992reducing_traffic, sorin2011primer}. 


An alternative approach is to equip the switch with persistence capability and treat it as an independent persistence structure. By construction, such a design can guarantee write propagation and write serialization, which are the primary objectives of coherence protocols. As illustrated in Fig.~\ref{fig:motivation_dpd}(a), when a write arrives at \(S1\), the data can be persisted immediately without issuing additional network requests or acquiring write permission in advance. However, a persistent CXL switch must be designed such that a subsequent read (\circled{3}) is either routed to S1, or S1 propagates the latest version to M3 so that M3 also reflects the most recent data. Despite this opportunity, there is currently no formalized framework for determining the additional design requirements needed to ensure that such a persistence domain preserves correctness and crash consistency. Therefore, it is essential to formally define this persistence domain and establish its correctness properties in order to derive the design criteria for persistent CXL switches that maintain correctness and crash consistency as the system scales. In this work, we leverage the CXL network design to enforce write serialization and propagation, thereby preserving the order of write operations until they reach the memory controllers. This further ensures that each read operation observes the latest version of the data. In addition, we explore software-based draining mechanisms to handle changes in the set of persistence structures which will be discussed in what follows.

\vspace{-1em}
\subsection{System Assumption}
To determine PCS design criteria, we first need to determine the system architecture, as system topology and routing mechanism significantly impact correctness. As a baseline system, we assume a disaggregated memory system in which multiple compute nodes (i.e., hosts) are connected to multiple memory nodes through a network of CXL switches, as illustrated in Fig.~\ref{fig:system_design}(a). The CXL fabric is organized as a simple 2x2 fully connected topology with deterministic routing~\cite{sullivan2022cxl_fabric_introduction}. This topology achieves low latency at the expense of  bandwidth, as latency is the critical performance factor for persistent programming. Each compute node includes local DRAM, while remote persistent memory nodes are attached via CXL as part of a disaggregated memory pool.  Programs use the \textit{clflush} instruction to evict cache-line copies from the volatile domain to the persistence domain for durability, and \textit{mfence} to enforce the persist barrier. For the CXL switch design, although no standardized publicly available hardware implementation currently exists, CXL technology builds upon the PCIe standard~\cite{cxl_specification2024}. Accordingly, our CXL switch design adheres to both CXL and PCIe specifications, and its latency characteristics are consistent with prior work~\cite{pond_2023}.

\begin{figure*}[!h]
    \centering
    \includegraphics[width=0.95\textwidth]{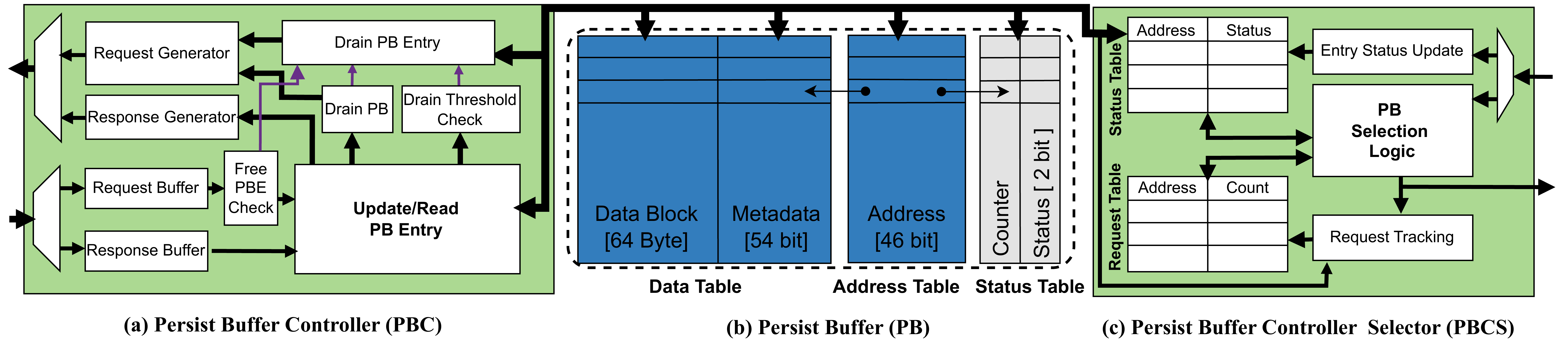}
    \caption{ Hardware component of persist buffer design. PBC Selector (c) resides in the main control logic, while Persist Buffer (b) and Persist Buffer Controller (a) are separate structures, outside the critical path of CXL switch operation. }
        \vspace{-1em}
    \label{fig:pb_controller_design}
\end{figure*}


 \section{Persistent CXL Switch Design}
\subsection{Adding Persist Capability at CXL Switch}
\label{label:pcs_design}

Our proposed CXL switch extends conventional routing functionality with persistency support. Specifically, the proposed switch enables (1) switch-level data persistence to reduce persist latency and (2) read forwarding to lower read latency. To provide these capabilities, the CXL switch design includes three components: (1) persist buffers (PB), (2) a persist buffer controller (PBC), and (3) a persist buffer controller selector (PBCS), as shown in Fig.~\ref{fig:system_design}(b). These switch modifications are integrated without adding delay to the critical path of the switch.



\subsubsection{Persist Buffer (PB)}

The proposed Persistent Buffer (PB) is organized as a fully associative structure that enables data persistence at the switch level. The PB may utilize non-volatile memory cells such as STT-RAM or volatile memory cells such as SRAM. In the latter case, we assume UPS or battery backing that can drain PB to the persistent memory in the event of a CXL power loss. 

The PB is organized into three tables: the \textit{Data Table}, the \textit{Address Table}, and the \textit{Status Table}. Each PB entry (PBE) consists of a single \textit{Address} entry in the \textit{Address Table}, which points to corresponding entries in both the \textit{Data Table} and \textit{Status Table}. This separation allows address lookup, data lookup, and status updates to proceed independently, reducing critical latency. The \textit{Status Table} is maintained as volatile storage because its information is not required for crash recovery.

As illustrated in Fig.~\ref{fig:pb_controller_design}(b), each \textit{Data Table} entry stores a 64-byte data block along with 54 bits of metadata (e.g., SPID, DPID, Opcode, LD-ID), which is necessary to reconstruct a CXL packet for forwarding updates to the next persistent structure. Each \textit{Address Table} entry maintains the 46-bit physical address (CXL packets include only bits [51:6] of the physical address). The \textit{Status Table} contains a 2-bit status field and an LRU counter (5 bits for 32 PBEs).

Each PB entry has one of four status values: \textit{Data}, \textit{Drain Issued}, \textit{Drain}, and \textit{Free}. \textit{Data} indicates that the PBE holds the newest version of the data and no downstream persistent structure (e.g., PCS or PM) has this version or a newer one. \textit{Drain Issued} indicates that the data has been issued to the next persistent structure but has not yet left the PBC’s port. \textit{Drain} indicates that the drain packet has been routed to the output port and the PBC is awaiting an acknowledgment confirming persistence at the next structure. \textit{Free} indicates that a newer copy has been persisted downstream and the entry may be reclaimed.

\subsubsection{Persist Buffer Controller (PBC)}

The Persist Buffer Controller (PBC) is the primary control logic in our PCS design. As shown in Fig.~\ref{fig:system_design}(b), the PBC is connected to an additional port (the PB port), enabling the PCS to redirect write requests to the PB without complicating the routing of regular CXL traffic (e.g., CXL.cache, CXL.io).

The PBC consists of several logic blocks, illustrated in Fig.~\ref{fig:pb_controller_design}(a). The \textit{Request Buffer} and \textit{Response Buffer} are FIFO buffers that temporarily hold incoming requests and responses, respectively. The PBC prioritizes the \textit{Response Buffer} to ensure that write acknowledgments are processed first, allowing associated PBEs to transition to the \textit{Free} state earlier.

The \textit{Update/Read PB Entry} block is responsible for updating PBEs for incoming write requests, serving read requests directly from the PB when data is present, and updating PBE and Status Table entries to the \textit{Free} state once persistence is confirmed. The \textit{Free PBE Check} logic verifies that a free PBE is available before forwarding a new write request from the \textit{Request Buffer}. If no \textit{Free} PBE exists, it triggers PBE draining (purple arrow in Fig.~\ref{fig:pb_controller_design}(a)), ensuring that new write requests will eventually be serviced.

The \textit{Drain PB Entry} block manages the draining of \textit{Data} PBEs. Draining is governed by a drain threshold (DT), which is set below the total number of PBEs. When the number of PBEs in the \textit{Data} state exceeds DT, the PCS begins draining entries according to an LRU policy until the number of \textit{Data} PBEs falls to DT or below.

\subsubsection{Persist Buffer Controller Selector (PBCS)} 
\label{label:selector_design}

The Persist Buffer Controller Selector (PBCS) is a logic block integrated into the CXL switch’s control logic. The control logic is the component that determines routing for incoming packets. When the control logic receives a CXL packet, it forwards the packet’s address, request type, and requesting-port information to the PBCS. This allows the PBCS to operate in parallel with the control logic, which continues performing its standard operations (e.g., routing-table lookup, virtual-channel allocation). 


The PBCS contains two tables: the \textit{Request Table} and a \textit{Status Table}. The \textit{Request Table} tracks all write requests that have been routed to the PBC but not yet processed by the \textit{Update/Read PB Entry} block. The \textit{Status Table} at the PBCS keeps a copies of the PBC \textit{Status Table} entries to enables fast queries of PBE status; since the PB is an independent component of the traditional CXL switch design, directly accessing PBC \textit{Status Table} could extend the critical path of the CXL controller. The local \textit{Status Table} in PBCS avoids this issue and allows timely selection decisions.

The \textit{PB Selection Logic} uses the \textit{Status Table} and \textit{Request Table} to determine whether an incoming read or write request should be routed to the PBC. This decision is then forwarded to the CXL controller to finalize routing. When a write request is routed to the PBC, the \textit{Request Tracking} unit updates the \textit{Request Table} to record the addresses and count the outstanding writes for that addresses, as multiple writes may target the same location.

When a packet arrives from the PB port, the \textit{Entry Status Update} unit updates the corresponding PBE status from \textit{Drain Issued} to \textit{Drain}. Collectively, the \textit{Status Table} and \textit{Request Table} ensure that the system maintains the required \textit{write propagation } and \textit{write serialization} correctness guarantee.

\subsection{CXL Switch Persist Working Principle}
\subsubsection{\textbf{Write/Read Request Processing}}

CXL controller sends only two request to PBCS : \textsl{Read Request}, \textsl{Write Request}. For any \textsl{Read} request from the PB port (Fig. \ref{fig:system_design}(c)), control logic does not send any inquiry to PBCS and instead route to destination port. 

For a write request \textsl{Write A}, the PBCS first checks whether address \textsl{A} already exists in the \textit{Status Table} or \textit{Request Table}. If so, it indicates the control logic to route the request to the PBC to satisfy persistent CXL switch (PCS) correctness requirements. Otherwise, PBCS evaluates the number of entries with \textit{Data} status in the \textit{Status Table} plus the outstanding write count in the \textit{Request Table}. If this total exceeds the PB capacity, the write is forwarded to the next persistent structure, as the PBC lacks available \textit{Free}, \textit{Drain}, or \textit{Drain Issued} PBEs and would require an additional drain to accommodate \textsl{Write A}. If capacity is sufficient, the PBCS routes \textsl{Write A} to the PBC and records it in the \textit{Request Table}. 


For a read request \textsl{Read A}, the PBCS checks whether \textsl{A} exists in the PB with \textit{Data} or \textit{Drain Issued} status in the \textit{Status Table} or has a pending entry in the \textit{Request Table}. If found, the read is routed to the PBC. The \textit{Request Table} enables tracking of in-flight \textsl{Write A} operations before the \textit{Status Table} is updated, ensuring correct write–read ordering.

When a request (\textsl{Write} or \textsl{Read}) enters the PBC, it is first inserted into the \textit{Request Buffer} and then forwarded to the \textit{Update/Read PB Entry} unit. The \textit{Response Buffer} is given higher priority than the \textit{Request Buffer} when issuing packets to the \textsl{Update/Read PB Entry}. For a write request, the \textit{Free PBE Check} ensures that either an existing entry with address A (an older version) or at least one \textit{Free} PBE is available in the PB before \textit{Update/Read PB Entry} begins processing the \textsl{Write}. Without this, a situation may arise where \textit{Update/Read PB Entry} dequeues a \textsl{Write} from the \textit{Request Buffer} but cannot find a \textit{Free} PBE, preventing it from handling subsequent \textsl{Acknowledgment} packets—potentially causing a deadlock. The \textit{Free PBE Check} therefore guarantees that, even in the worst case, \textit{Update/Read PB Entry} remains able to process future response packets. Once \textit{Update/Read PB Entry} begins processing \textsl{Write A}, it extracts the data and metadata from the packet and writes them into the PB, either overwriting the existing entry or allocating a new PBE. It then updates the status to \textit{Data} in the \textit{Status Table} at both the PB and PBCS, and refreshes the LRU counter. After the persist operation completes, the unit sends \textsl{Write A} to the \textit{Response Generator}, which produces a \textsl{Write-acknowledgment} and returns it to the requester. In parallel, the \textit{Drain Threshold Check} monitors the number of \textit{Data} PBEs; if this count exceeds the drain threshold (DT), a single \textit{Drain PB Entry} operation is triggered to drain an existing PBE.

For a \textsl{Read} request, the \textit{Update/Read PB Entry} unit first checks whether the requested data exists in the PB. If the data exists, it retrieves the data from the PB and forwards both the data and \textsl{Read A} to the \textit{Response Generator}, which converts the request into a response containing the data and returns it to the requester. In the rare case where the data is not found (e.g., entry A has already been drained before \textsl{Read A} reaches \textit{Update/Read PB Entry}), \textsl{Read A} is forwarded to the \textit{Response Generator}, which simply routes the request to the next structure along the persistent path. When a \textsl{DrainPath} (OS issued drain request; see \ref{label:drain_path}) request is received, \textit{Update/Read PB Entry} sends it to \textit{Drain PB}, which temporarily holds the request packet and repeatedly signals \textit{Drain PB Entry} to drain all \textit{Data} PBEs. After draining completes, the \textsl{DrainPath} request is forwarded to the next structure via the \textit{Request Generator}, which does not modify the packet but simply places it into the output buffer.

\subsubsection{\textbf{Draining Persistent Buffer Entry}}
\label{drain_request_process}

When the \textit{Drain PB Entry} receives a drain signal from either \textit{Free PBE Check}, \textit{Drain PB}, or \textit{Drain Threshold Check}, it selects a \textit{Data} PBE according to the replacement policy (e.g., LRU), updates its status to \textit{Drain Issued} in both the PB and PBCS \textit{Status Table}, retrieves the corresponding data and metadata from the PB’s \textit{Data Table}, and sends this information to the \textit{Request Generator}. The \textit{Request Generator} reconstructs a \textsl{Write A} request and issues it to the PBC-attached port for routing along the persist path.

During packet processing, the control logic forwards the request to the PBCS. Upon receiving this regenerated \textsl{Write A}, the PBCS instructs the control logic to route it to the appropriate destination and updates the \textit{Status Table} entry from \textit{Drain Issued} to \textit{Drain}. Because this update is off the critical path of the control logic, it does not introduce performance bottlenecks for normal packet routing.

When the control logic encounters a \textsl{Write-acknowledge A} packet from downstream switches, it forwards the packet to the PBCS, which consults the \textit{Status Table} for the corresponding entry. If a \textit{Drain} PBE exists for address A, the packet is routed to the PBC. Once the PBC receives \textsl{Write-acknowledge A}, it places the packet in the \textit{Response Buffer}. Eventually, the \textit{Update/Read PB Entry} unit processes the \textsl{Write-acknowledge A} and updates the PBE status from \textit{Drain} to \textit{Free}.

\section{Correctness Assurance for Distributed Persistence Domain}

Ensuring correctness is essential in persistent memory disaggregation, particularly when persistence is shifted into the network. This approach deviates from the traditional notion of persistent memory by introducing persistence capabilities at intermediate stages between compute nodes and the persistent memory devices. 

For correctness, the distributed persistent domain  has to ensure:
    \begin{enumerate}[label=\alph*)]
     \item A read request must return the most up-to-date version of the data.
    \item The DPD must preserve the latest version of the data without loss or unintended overwriting.
\end{enumerate}

To satisfy the correctness assumptions, our proposed design must fulfill the following requirements: (1) preserving the ordering among different versions of the same data, (2) ensuring that each read request reaches the latest copy of the data, (3) at least one copy of the latest data is within the persistent domain, and (4) supporting data drain operation when prior three requirements cannot be fulfilled.



\subsection{PCS Design for Ensuring Correctness}
\label{label:dpd_correctness_check}

As shown in Fig.~\ref{fig:pb_controller_design}(c), the PBCS maintains a \textit{Status Table} to track data that has already been persisted in the PB, and a \textit{Request Table} to record requests that have already been routed to the PBC. These structures enable the PCS to determine whether it currently holds a copy of a given data block and, if so, to route subsequent requests for that data to the PBC. For a write request, if the PBC finds an existing copy of the data in the PB, it overwrites that copy with the newly arrived version. This removes the older version from the PCS, thereby eliminating the possibility of write reordering and effectively guaranteeing write propagation. For a read request, if the PBC finds a copy in the PB, it responds using its local copy of the data, which is guaranteed to be the latest version.

When the PCS forwards data along the persist path to free a PBE, it keeps the corresponding PBE in the \textit{Drain} or \textit{Drain Issued} state until it receives a \textit{Write-acknowledgment} packet from the next persistence structure. This guarantees that another persistence structure has already stored a copy of the data, making it safe to mark the entry as \textit{Free}. When the PCS completes persistence of a data block, it also sends a \textit{Write-acknowledgment} to the requester so that the requester is aware that the data has been safely persisted.

\begin{figure}[!t]
    \centering
    \includegraphics[width=\linewidth]{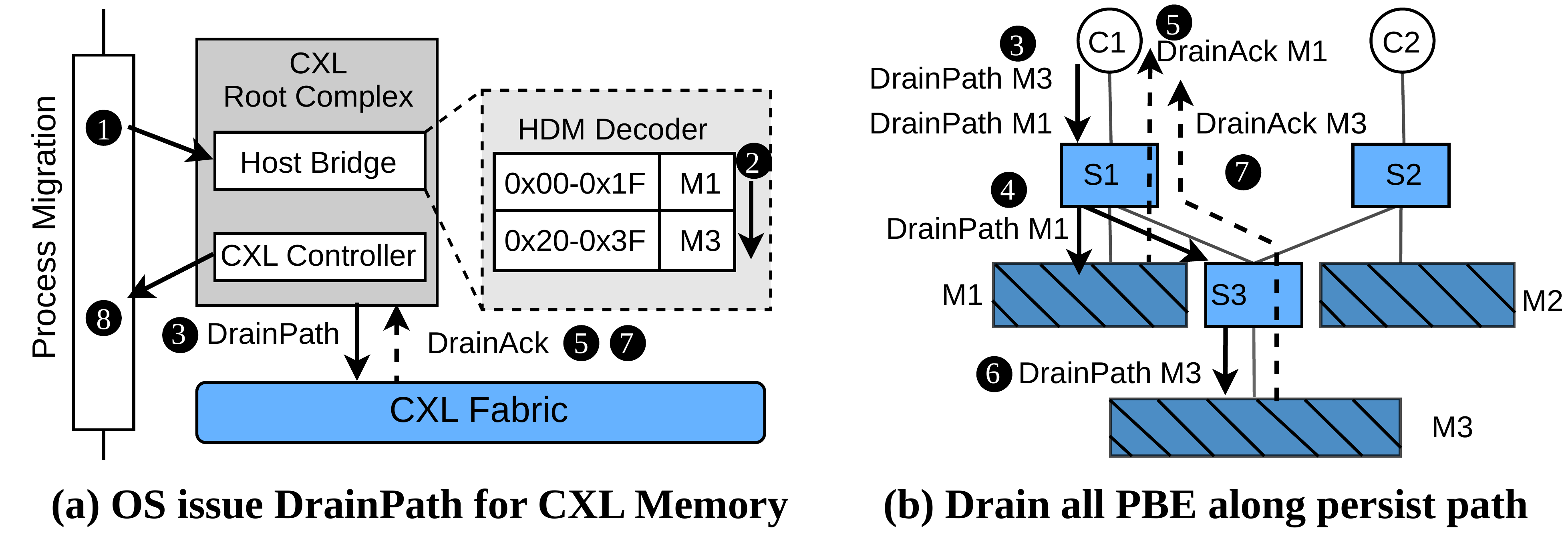}
    \vspace{-2.5em}
    \caption{When migrating a process to a new compute node, the OS issues \textsl{DrainPath} requests for all CXL-attached memory so that a new persistence path can be established.}
    \vspace{-1em}
    \label{fig:drain_path}
\end{figure}

\subsection{Software Support for Process Migration} 
\label{label:drain_path}



In a disaggregated system, multiple compute nodes connect to a memory pool through CXL switches to provide additional memory capacity on demand across all available compute nodes. This setup allows a process to be migrated to a different node for load balancing. As a result, data ownership may change when the operating system migrates a process between compute nodes. Such ownership changes could violate the correctness guarantees due to the changed persistent path. To prevent the potential correctness issues illustrated in Fig.~\ref{fig:motivation_dpd}, all data along the existing persistence path must be drained before ownership changes take effect. To this end, we propose a software mechanism, called the \textsl{DrainPath} operation, to enforce \textit{design requirement 1} and preserve correctness when data ownership changes.

Fig.~\ref{fig:drain_path} shows a \textsl{DrainPath} operation. During process migration, once the OS determines that the migrating process will not issue new writes until migration completes, it signals (\circled{1}) the CXL root complex to issue a \textsl{DrainPath} request for all mapped CXL memory. The root complex consults its HDM decoder to enumerate all attached CXL memory devices (\circled{2}) and sends a \textsl{DrainPath} request to each device (\circled{3}). When a PCS receives a \textsl{DrainPath} request (regardless of destination ID), it buffers the request in the \textit{Drain~PB} block (Fig.~\ref{fig:pb_controller_design}(a)) and instructs \textit{Drain~PB~Entry} to drain all PBEs until no \textsl{Data} PBE remains. Only after draining does PBC forward the \textsl{DrainPath} request to the next PS, eventually reaching the destination memory node (\circled{4}). Because each PCS guarantees that all \textsl{Data} PBEs are drained before forwarding, the destination memory node has already received all in-flight writes by the time it observes the \textsl{DrainPath} request. Thus, upon receiving \textsl{DrainPath}, the memory node immediately returns a \textsl{DrainAck} to the requesting compute node (\circled{5}).

With memory pooling, different memory nodes may receive the \textsl{DrainPath} request at different times (\circled{4}, \circled{6}), and their corresponding \textsl{DrainAck} responses may therefore arrive at different times as well (\circled{5}, \circled{7}). Once all \textsl{DrainAck} messages have been collected, the CXL root complex notifies the OS (\circled{8}) that migration can safely complete, after which the migrated process may resume memory accesses.


\subsection{Crash Recovery}

\subsubsection{Recovery from processor-side crash:} 
If any processor-side component fails (i.e., a crash occurs within one or more compute nodes), we assume that the process has a properly implemented crash-recovery mechanism and that the system provides a way to remap device physical addresses to operating physical addresses to support recovery~\cite{fiala2016mini_ckpts, assa2026programming_model, oliveira2025rethinking_pm_crash, guo2026cxlmc}. In this case, our DPD imposes no additional correctness constraints, as long as the process is not migrated to a different compute node after reboot. However, if the process is resumed on a different compute node after reboot, the operating system issues \textsl{DrainPath} requests to all persistent memory modules attached to the DPD. If the crash leaves the OS without complete knowledge of the attached memory nodes, it can issue the \textsl{DrainPath} requests to all switches. This process guarantees that all versions of the data are drained to the remote memory nodes. Consequently, each memory node holds the latest version of the data, and any future read request will be routed to the correct node and retrieve the most up-to-date value.

\subsubsection{Recovery from memory-side crash:}
On the memory side, crashes may be partial (e.g., a single switch fails and reboots) or complete (e.g., the entire CXL fabric crashes and all components reboot). If such a crash occurs, our PCS design preserves all data copies residing in the PB. After a full reboot, however, multiple versions of the same data may exist across different PCS units in the DPD. Correct recovery therefore depends on two conditions: (1) accurately identifying the most recent version among all recovered copies, and (2) ensuring that all subsequent reads return this latest version. The DPD system can achieve this using the \textsl{DrainPath} mechanism. Upon recovery from a memory-side reboot, the CXL fabric manager itself (or the OS at its request) issues \textsl{DrainPath} commands to drain all copies to the remote memory node.


{

\subsubsection{Fault tolerance}

In a CXL memory fabric with volatile CXL switches, the primary point of hardware failure is the persistent memory device, which can lead to permanent data loss. To prevent catastrophic loss, systems may employ various fault-tolerance mechanisms, such as module-level redundancy~\cite{krishnan2023tenet, kateja2020tvarak} or mirroring~\cite{tavakkol2018rdma_mirroring_pm, wang2023rowan, zheng2015ASPLOS_mojim}. In our DPD design, we assume that PCS fails together with the corresponding memory module. Upon fault detection, the fabric manager uses the \textsl{DrainPath} mechanism to drain all PB entries within the DPD, ensuring that all copies reside in remote memory nodes, and then initiates backup-data recovery as if the memory module had encountered an uncorrectable media error~\cite{meza2015flash_error_study}

}


\begin{table}[t!]
\centering
\caption{Simulation Environment}
\vspace{-1em}
\begin{tabularx}{\linewidth}{|l|>{\arraybackslash}X|}
\hline

    Processor    & 8-core OoO SimpleSwitchableProcessor\\ \cline{2-2}
                & 4 GHz, 64-bit X86 ISA \\
    
    \hline
    \multirow{4}*{Cache}& 64B block, inclusive, LRU replacement policy \\ \cline{2-2}
                        & 32KB, 8-way associative private L1 cache\\ \cline{2-2}
                        &  8KB , 4-way associative TLB cache \\ \cline{2-2}
                        & 256KB, 8-way associative Shared L2 cache  \\
    \hline
    \multirow{4}*{Memory}    & Total Memory pool capacity: 32 GB \\\cline{2-2}
                            &   DRAM : DIMM\_DDR5\_4400  \\\cline{2-2}
                                & PM read latency: 150ns  \\\cline{2-2} 
                                & PM write latency: 500ns \\
    \hline \hline
    \multirow{2}*{CXL switch}  & 16  port, x16 lane per port \\\cline{2-2}
                            &  128GB/s per port, 68Byte flit\\
    \hline

    \multirow{3}*{Persist Buffer}  & 32 entry, LRU replacement policy\\\cline{2-2}
                            & 0.295 ns data access latency \\

    \hline
\end{tabularx}
\label{tab:hardware_parameter}
\end{table}

\section{Experimental Setup}

\subsection{Simulation Environment}

We evaluated our distributed persistence domain design using gem5 v23.1~\cite{gem52020, gem52011}, an event-driven simulator with parameters shown in Table~\ref{tab:hardware_parameter}. We used CACTI~\cite{cacti1996} to determine the access latency of the persistent buffer in the PCS. Persistent memory read and write latencies were set to 150 ns and 500 ns, respectively, which are commonly assumed values in persistent memory system simulations ~\cite{alshboul-hpca21-bbb, hadi2023DAI} and consistent with prior studies~\cite{chatzistergiou2015rewind, kim2016nvwal}. For the CXL switch design, we used a PCIe switch architecture as a reference and adopted a 68-byte flit size, which is supported across all versions of the CXL protocol~\cite{cxl_specification2024}. We incorporated POND~\cite{pond_2023} to model timing behavior for our CXL switch implementation. In our initial evaluation, we used 32 persistent buffer entries per CXL switch following prior work~\cite{alshboul-hpca21-bbb, freij2023secpb}. For the main evaluation, we assumed a single-hop distance between the compute node and memory node, as operating systems in disaggregated systems typically allocate new memory requests to the closest memory node.

\subsection{Evaluated Workload}
\label{label:workload}

\begin{table}[t!]
\centering
\caption{Splash-4 Benchmark Suite}
\vspace{-1em}
\begin{tabularx}{\linewidth}{|>{\centering\arraybackslash}X|>{\centering\arraybackslash}X|}
\hline
    \rowcolor{lightgray} Benchmark & Input \\
    \hline \hline
    BARNES          &  n8192            \\ \hline
    CHOLESKY          &  tk23.O        \\ \hline
    FFT                 &  -m16  -l6  -n128        \\ \hline 
    LU\_CONT        &  -n256 -b50            \\ \hline 
    LU\_NON             & -n128 -b50         \\ \hline
    RAYTRACE          & teapot.env        \\ \hline
    VOLREND      & head-scaleddown2       \\ \hline  
    VOLREND\_NPL    & head-scaleddown2      \\ \hline
\end{tabularx}
\label{tab:splash}
\vspace{-1.1em}
\end{table}

\begin{table}[t!]
\centering
\caption{YCSB Benchmark Suite}
\vspace{-1em}
\begin{tabularx}{\linewidth}{|c|c|>{\centering\arraybackslash}X|}
\hline
    \rowcolor{lightgray} Benchmark & Type & Composition \\
    \hline \hline
    A          &  Write heavy   &  50\% Read, 50\% Update       \\ \hline
    B          &  Read mostly   &   95\% Read, 5\% Update  \\ \hline
    C          & Read only  &     100\% Read   \\ \hline 
    D        &  Read-Insert &      95\% Read, 5\% Insert      \\ \hline 
    F             & Atomic &  50\% Read, 50\% RMW       \\ \hline
\end{tabularx}
\label{tab:ycsb}
\end{table}

\begin{figure*}[!t]
    \centering
    \includegraphics[width=0.95\textwidth]{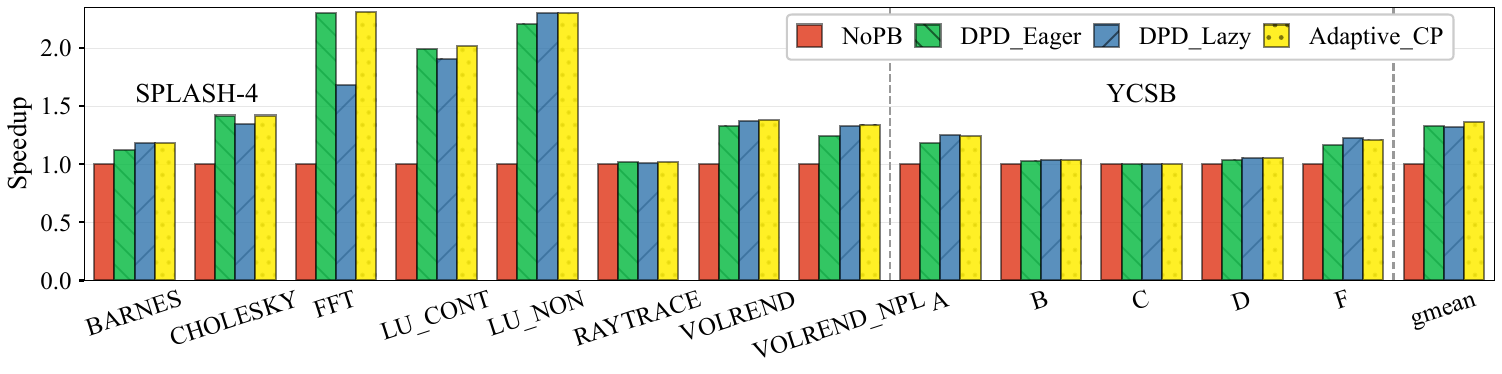}
        \vspace{-2em}
    \caption{Speedup of SPLASH-4 and YCSB benchmarks for different PCS schemes }
        \vspace{-1.5em}
    \label{fig:speedup}
\end{figure*}

To evaluate our work, we use the SPLASH-4~\cite{splash4} benchmark and the Yahoo! Cloud Serving Benchmark (YCSB) \cite{cooper2010ycsb} with memcached ~\cite{fitzpatrick2011memcached}. SPLASH-4 is a multi-threaded, compute-intensive workloads that can leverage persistent memory pooling for crash consistency and fast crash recovery \cite{fridman2021use_pm_hpc}. Memcached is a widely used key-value store database that could leverage memory pooling capability to serve memory demand beyond local memory capacity ~\cite{lim2012system_DM, liu2019mem_disagg_research}. We assume that all dynamically allocated memory (i.e., heap memory) is mapped to remote persistent memory. For SPLASH-4, we adopt efficient checkpointing with recomputation~\cite{efficient_checkpointing2017} to ensure crash consistency in the persistent memory system. For memcached, YCSB client requests are used as input, and after each Update and Insert operation, the updated data is persisted. To initialize the database, we choose a record count of 100,000 and run 10,000 operations for each workload, commonly chosen parameters for YCSB with key-value store database  in simulated environment ~\cite{ham2024MICRO_m2ndp, zeng2023persitent_process, colaso2017no_sql_mem_characterization}. Table~\ref{tab:splash} and ~\ref{tab:ycsb} list all workloads used in our evaluation. Workloads that do not execute across all schemes or are not supported by memcached are ignored. Each benchmark uses 8 threads. For each benchmark, we first run the program in atomic CPU mode to accelerate simulation until it reaches the parallel region of interest (ROI). We then switch to detailed simulation mode to collect statistics.
 
\subsection{Baselines}

For our base system, we assume a CXL-switch-based distributed persistent memory architecture without PB support (NoPB). For PCS, performance can be influenced by the drain threshold value, which determines when the PCS decides to drain data from the PCS to the next persistent structure. The drain threshold may be fixed or adaptive. For fixed drain thresholds, we evaluate two schemes: eager draining and lazy draining. In the DPD with eager draining (DPD\_Eager) scheme, the PCS effectively sets the drain threshold to zero. This ensures that immediately after data is persisted and the \textsl{Write-acknowledgment} is sent back to the requester, the PCS begins draining the data block. In the DPD with lazy draining (DPD\_Lazy) scheme, the PCS tries to retain a PBE for as long as possible by setting a high drain threshold and only draining the data to PM when additional resources are needed. For our evaluation, we set the drain threshold to 75\% for DPD\_Lazy, a commonly used value for buffer-like hardware designs~\cite{alshboul-hpca21-bbb, freij2023secpb}.

We tested several schemes that try to adopt drain threshold to achieve the best performance for DPD\_Lazy and DPD\_Eager across all benchmarks and observe how different approaches to adapting DT impact performance. Detail exploration regarding different adaptive schemes are omitted due to page limitations.   For the main evaluation, we selected Adaptive\_CP, the best-performing scheme among all the adaptive schemes we tested, which tries to predict upcoming write burst based on PB utilization. Adaptive\_CP employs a simple adaptive mechanism that increase/decrease drain threshold value  by a constant (C) value if  available PBE utilization/availability  goes above/below a certain percentage (P). To avoid bias toward eager or lazy draining scheme, we set initial drain threshold value at 50\%.


\section{Evaluation}
\subsection{Performance Evaluation}

Fig.~\ref{fig:speedup} shows the speedup for different design schemes relative to the base system, NoPB. We observe that, on average, our scheme achieves a $32\% \sim 36\%$ speedup when compared to NoPB. Among compute heavy workloads (SPLASH-4),  FFT, LU\_CONT, \& LU\_NON achieve significant improvements (for example, FFT achieves a 131\% speedup under Adaptive\_CP), while benchmarks such as BARNES show more modest gains (18\% speedup for DPD\_Lazy). RAYTRACE is an exception and does not exhibit noticeable benefits from any scheme (less than 2\% speedup). For the database (YCSB+Memcached),  write heavy workload (A \& F)  shows modest performance gain (25\% \& 23\% respectively for DPD\_Lazy). Workload B \& D have only 5\% write operations, resulting in small performance gain (4\% \& 6\% respectively for DPD\_Lazy) where workload C has no impact (speedup or slowdown) as it is a read only workload. 

Across all  SPLASH-4 benchmarks, the adaptive scheme (Adaptive\_CP) consistently outperforms the fixed schemes (DPD\_Lazy \& DPD\_Eager). Across  all YCSB  benchmarks, Adaptive\_CP performs similar to DPD\_Lazy ($0\% \sim 2\%$ difference ) resulting in the best performing scheme across the evaluation (average 36\% speedup compared to 33\% and 32\% speedup of DPD\_Eager and DPD\_Lazy respectively). Understanding the underlying performance behavior of the fixed schemes helps explain why the adaptive schemes perform better overall. By design, DPD\_Eager focuses solely on reducing persist operation latency, whereas DPD\_Lazy also attempts to serve upcoming read requests at the CXL switch to reduce read latency.

We observe that for some benchmarks, DPD\_Eager outperforms DPD\_Lazy, particularly for FFT, where DPD\_Lazy achieves a 68\% speedup compared to DPD\_Eager's 130\% speedup. LU\_CONT and LU\_NON both achieve substantial speedups under DPD\_Eager (99\% and 121\%, respectively). However, under DPD\_Lazy, LU\_CONT loses some performance (reduced to 91\%), whereas LU\_NON gains additional improvement (increasing to 130\%).  For YCSB workloads , DPD\_Lazy outperform DPD\_Eager across workloads except C (read-only workload have near zero performance impact for PCS). However, no database workload has a performance gain similar to compute heavy workloads like FFT, LU\_CONT, LU\_NON.  

\begin{figure}[!t]
    \centering
    \includegraphics[width=\linewidth]{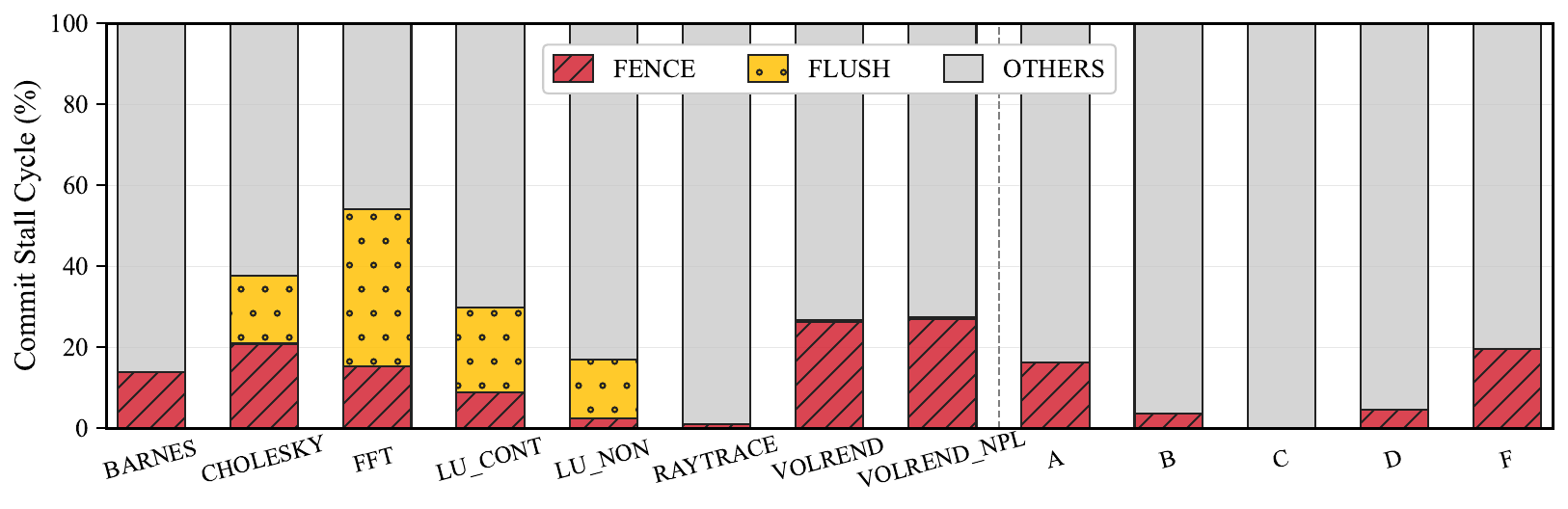}
    \vspace{-2.5em}
    \caption{Commit stall breakdown for SPLASH-4 and YCSB benchmarks in NoPB}
    \vspace{-1.5em}
    \label{fig:commit_stall_breakdown}
\end{figure}

\subsubsection{Commit Stall Breakdown}  Fig.~\ref{fig:commit_stall_breakdown} shows the commit-stage stall breakdown for the NoPB baseline. We separate stalls caused by persist operations (\texttt{FLUSH} and \texttt{FENCE}) from all remaining stalls (grouped as \textit{OTHERS}). RAYTRACE and C workloads exhibit negligible \texttt{FLUSH}/\texttt{FENCE}-induced stalls, explaining its minimal speedup. In contrast, FFT shows the highest fraction of persist-induced stalls and therefore achieves the largest performance gain through our design. However, under DPD\_Lazy, FFT loses most of this benefit. From the figure, we observe that FFT’s stalls are dominated by \texttt{FLUSH}, and its \texttt{FENCE}-related stalls are smaller than those of CHOLESKY, VOLREND, and VOLREND\_NPL. For all YCSB workloads, \texttt{FLUSH}  related stalls are nearly zero. 

\texttt{FLUSH}-related commit stalls occur when the L1 MSHR cannot accept new \texttt{FLUSH} requests, which can happen under bursty flush behavior. In DPD\_Lazy, only 75\% of PBEs are reserved to retain data for potential read forwarding. If a large number of \texttt{FLUSH} requests arrive simultaneously, the PCS may fail to drain PBEs fast enough, forcing writes to bypass the switch and causing PB write failures. Notably, BARNES, VOLREND, VOLREND\_NPL  show no \texttt{FLUSH}-related stalls. Each memcached database thread processes one record at a time thus there is no scenario where multiple \texttt{FLUSH} requests could overflow L1 MSHR, resulting zero \texttt{FLUSH} related stall. For these workloads, DPD\_Lazy outperforms DPD\_Eager.

\begin{figure}[!t]
    \centering
    \includegraphics[width=\linewidth]{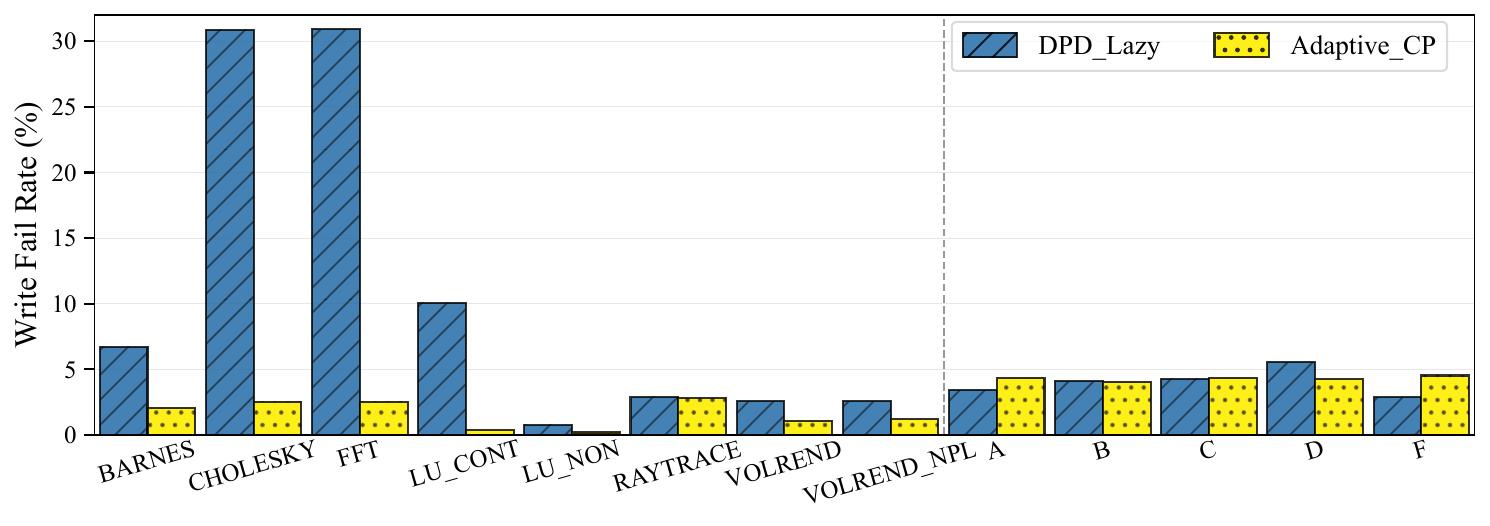}
        \vspace{-2.5em}
    \caption{Percentage of write request fail to persist  at PCS }
    \vspace{-1.5em}
    \label{fig:write_fail_rate}
\end{figure}

\begin{figure}[!t]
    \centering
    \includegraphics[width=\linewidth]{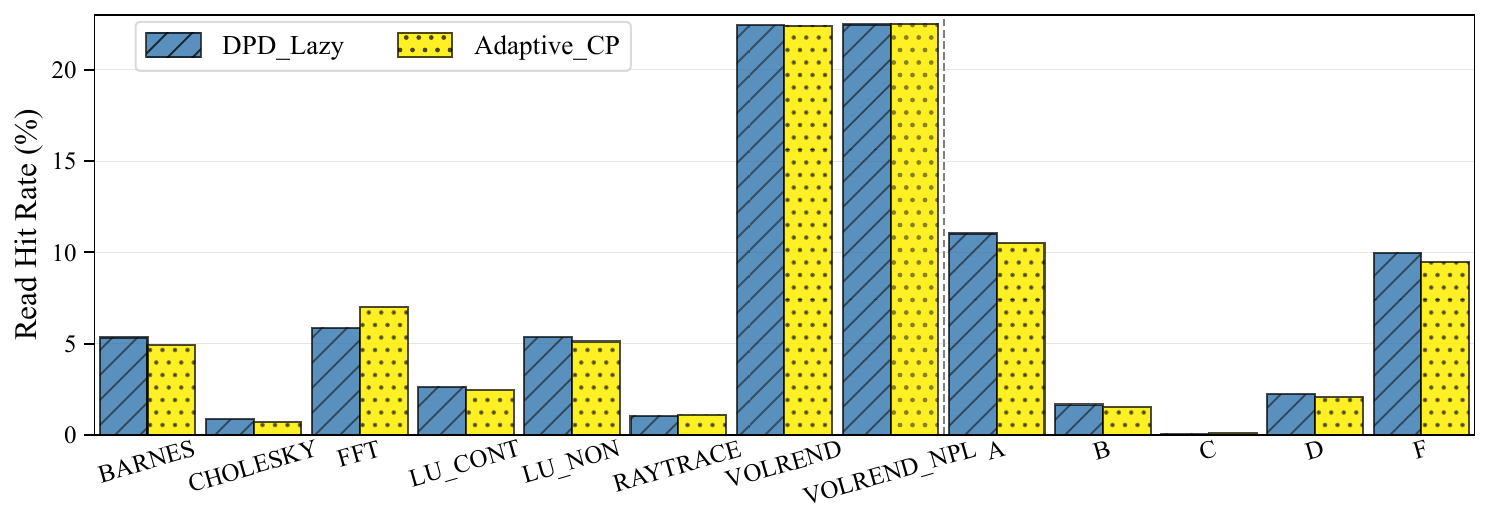}
        \vspace{-2.5em}
    \caption{Percentage of read request hit at PCS.}
    \vspace{-1.5em}
    \label{fig:read_hit_rate}
\end{figure}

\subsubsection{Write Failure Rate Analysis} Fig.~\ref{fig:write_fail_rate} shows the percentage of write requests that fail to persist at the PCS. DPD\_Eager is omitted because its write failure rate is minimal across all workloads, with the worst case being only 0.26\% for workload C. DPD\_Lazy exhibits a higher write failure rate than the other schemes because it reserves 75\% of the PBEs for holding data blocks. For FFT, CHOLESKY, and LU\_CONT, the write failure rate exceeds 10\%, and all three benchmarks perform worse under DPD\_Lazy than under DPD\_Eager. In particular, the 30\% failure rate of FFT explains its significant performance degradation relative to DPD\_Eager. Although CHOLESKY experiences a similar failure rate, its performance is not affected as severely because, as shown in Fig.~\ref{fig:commit_stall_breakdown}, CHOLESKY has far fewer \texttt{FLUSH}-related stalls than FFT, with most of its stalls instead caused by \texttt{FENCE}. Adaptive\_CP achieves lower write failure rates across all SPLASH benchmarks.

For YCSB, the write failure rate is below 6\% across all workloads; therefore, DPD\_Lazy delivers better performance than DPD\_Eager. For workloads A and F, Adaptive\_CP exhibits a slightly higher write failure rate than DPD\_Lazy (less than 2\%), which explains why Adaptive\_CP does not outperform DPD\_Lazy as it does for the SPLASH-4 workloads. In Fig.~\ref{fig:write_fail_rate}, we observe noticeable write failure rates for workloads B, C, and D, even though they are read-heavy or read-only workloads and do not show a significant performance impact on PCS. This behavior arises because Memcached maintains various runtime metadata, in addition to key-value information, that are updated during read operations. These metadata are also mapped to persistent memory as part of the data structure, but they are not critical for crash consistency and therefore are not persisted after updates, resulting in no performance impact. When multiple cache blocks are evicted, updates to this runtime metadata trigger multiple write-back requests to persistent memory, which is the source of the write failures in workloads B, C, and D shown in Fig.~\ref{fig:write_fail_rate}. However, because these write-backs do not stall program execution, PCS shows no noticeable performance degradation. These unintended write-backs from metadata updates also interfere with Adaptive\_CP's ability to adjust the drain threshold and optimize the performance trade-off between DPD\_Eager and DPD\_Lazy. As a result, Adaptive\_CP shows a higher write failure rate than DPD\_Lazy for workloads A and F, as observed in Fig.~\ref{fig:write_fail_rate}.

\subsubsection{Read Hit Rate Analysis}  Fig.~\ref{fig:read_hit_rate} shows the percentage of read requests that hit in the PCS, allowing the PCS to return data directly from the CXL switches and avoid a remote fetch from persistent memory, thereby reducing read latency. DPD\_Lazy attempts to retain data blocks longer so that subsequent reads can be served directly from the PB. For most benchmarks, the read-hit rate is below 7\%, except for VOLREND and VOLREND\_NPL in SPLASH-4, which achieve over 20\%, and workloads A and F in YCSB, which achieve approximately 10\%. We also observe that although BARNES and LU\_NON have read-hit rates of only about 5\%, they still achieve better performance under DPD\_Lazy than under DPD\_Eager. FFT exhibits a similar hit rate, but because of its very high write failure rate (Fig.~\ref{fig:write_fail_rate}), this benefit becomes negligible. Although our adaptive scheme does not always achieve the highest read-hit rate, its hit rate remains close to the best case (within 0.52\%). By maintaining a comparable read-hit rate while significantly reducing the write failure rate (Fig.~\ref{fig:write_fail_rate}), the adaptive scheme achieves the same or better performance than the fixed-threshold schemes (DPD\_Eager and DPD\_Lazy) for most benchmarks, except workloads A and F, where DPD\_Lazy outperforms Adaptive\_CP by about 1\%. For DPD\_Eager, in rare cases, a read request may arrive at the PCS while it is still draining data, resulting in a read hit. Across all workloads, however, the read-hit rate of DPD\_Eager is below 0.01\% and is therefore omitted from Fig.~\ref{fig:read_hit_rate}.


\begin{figure}[!t]
    \centering
    \includegraphics[width=\linewidth]{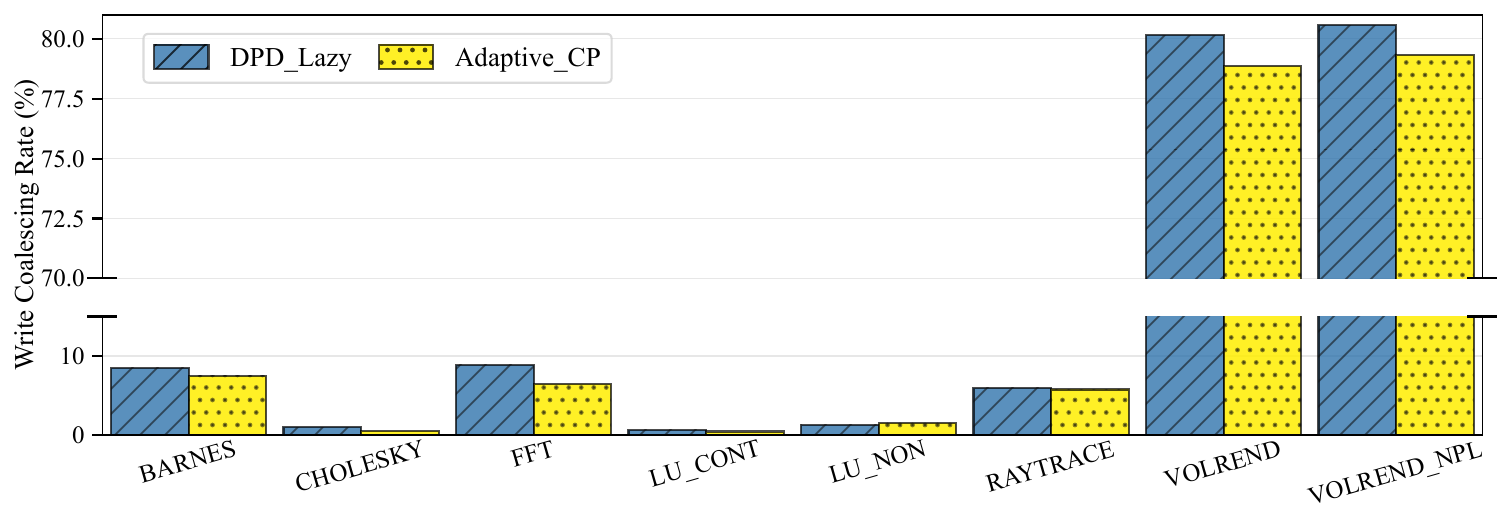}
    \vspace{-2.5em}
    \caption{Percentage of write coalesce at PCS.} 
    \vspace{-1.5em}
    \label{fig:write_coalesce}
\end{figure}

\begin{figure}[!t]
    \centering
    \includegraphics[width=\linewidth]{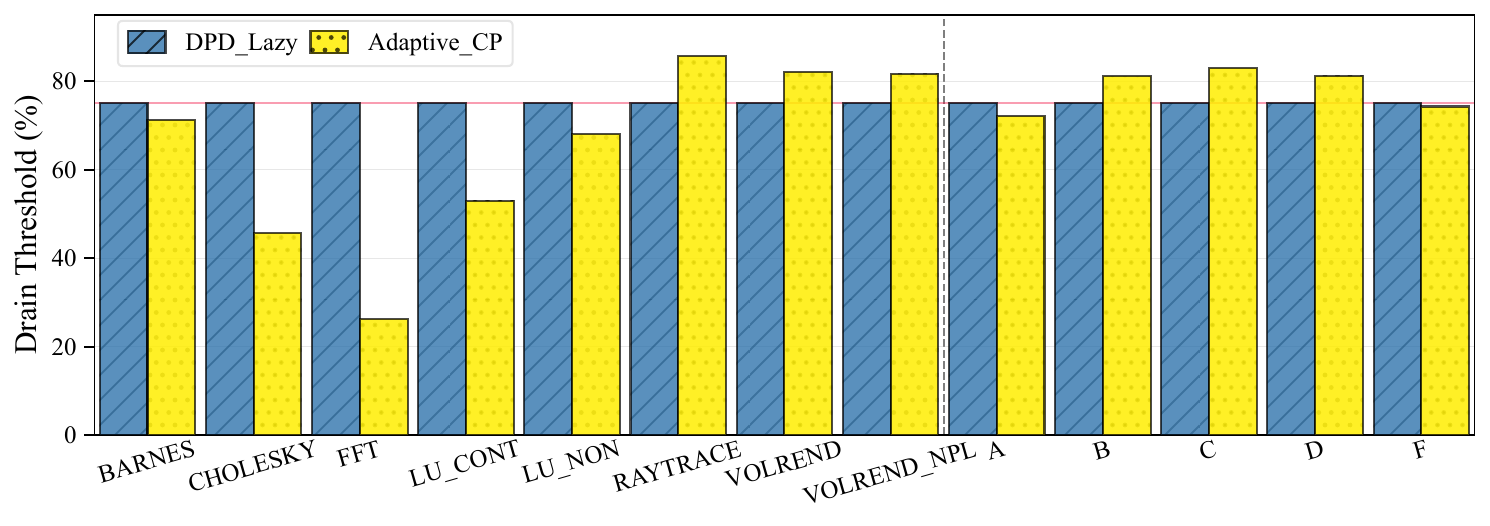}
    \vspace{-2.5em}
    \caption{Average drain threshold across different scheme. }
    \vspace{-1.5em}
    \label{fig:average_drain_threshold}
\end{figure}

 \subsubsection{Write Coalesce Analysis} Fig.~\ref{fig:write_coalesce} shows the percentage of write requests that are coalesced with an existing data block in the PB, which represents another potential benefit of using a higher drain threshold. We omit YCSB in Fig. \ref{fig:write_coalesce} as it shows less than 0.1\% write coalescing across all YCSB workloads. For SPLASH-4, overall write-coalescing rate follows a similar trend to the read-hit rate, as both metrics are closely tied to memory locality in persistent operations. For all SPLASH-4 benchmarks except VOLREND and VOLREND\_NPL, the write-coalescing rate is only 1–2\% higher than the read-hit rate. In contrast, VOLREND and VOLREND\_NPL achieve an over 80\% write-coalescing rate under DPD\_Lazy, meaning that only 20\% of CPU-issued writes are sent to persistent memory. Reducing the number of writes reaching PM would improve PM endurance, as persistent memory technology has limited write cycles. For Adaptive\_CP, the coalescing rate decreases by only 2–3\%, which is negligible. Although write coalescing does not significantly affect performance in our evaluated workloads, bandwidth-bound applications would benefit from issuing fewer downstream requests (via both write coalescing and read forwarding).

 \subsubsection{Draining Threshold Analysis} Fig.~\ref{fig:average_drain_threshold} shows the average drain threshold for Adaptive\_CP schemes across evaluated benchmarks (DPD\_Eager is omitted because its threshold is always 0\%; DPD\_Lazy is fixed at 75\% but included as a reference). For benchmarks where DPD\_Eager outperforms DPD\_Lazy (CHOLESKY, FFT, \& LU\_CONT), the adaptive scheme reduce the drain threshold  to lower the probability of write failures. Conversely, for benchmarks where DPD\_Lazy performs better than DPD\_Eager (BARNES, LU\_NON, VOLREND, VOLREND\_NPL and all YCSB workloads), the adaptive schemes increase the threshold above the initial 50\% setting to retain more data in the PBEs—sometimes even exceeding DPD\_Lazy’s 75\% threshold (e.g., RAYTRACE, VOLREND, and VOLREND\_NPL).

\subsection{Sensitivity Study}

\subsubsection{Persistent buffer size sensitivity}
\label{lab:pb_sensitivity}
\begin{figure}[!t]
    \centering
    \includegraphics[width=.48\textwidth]{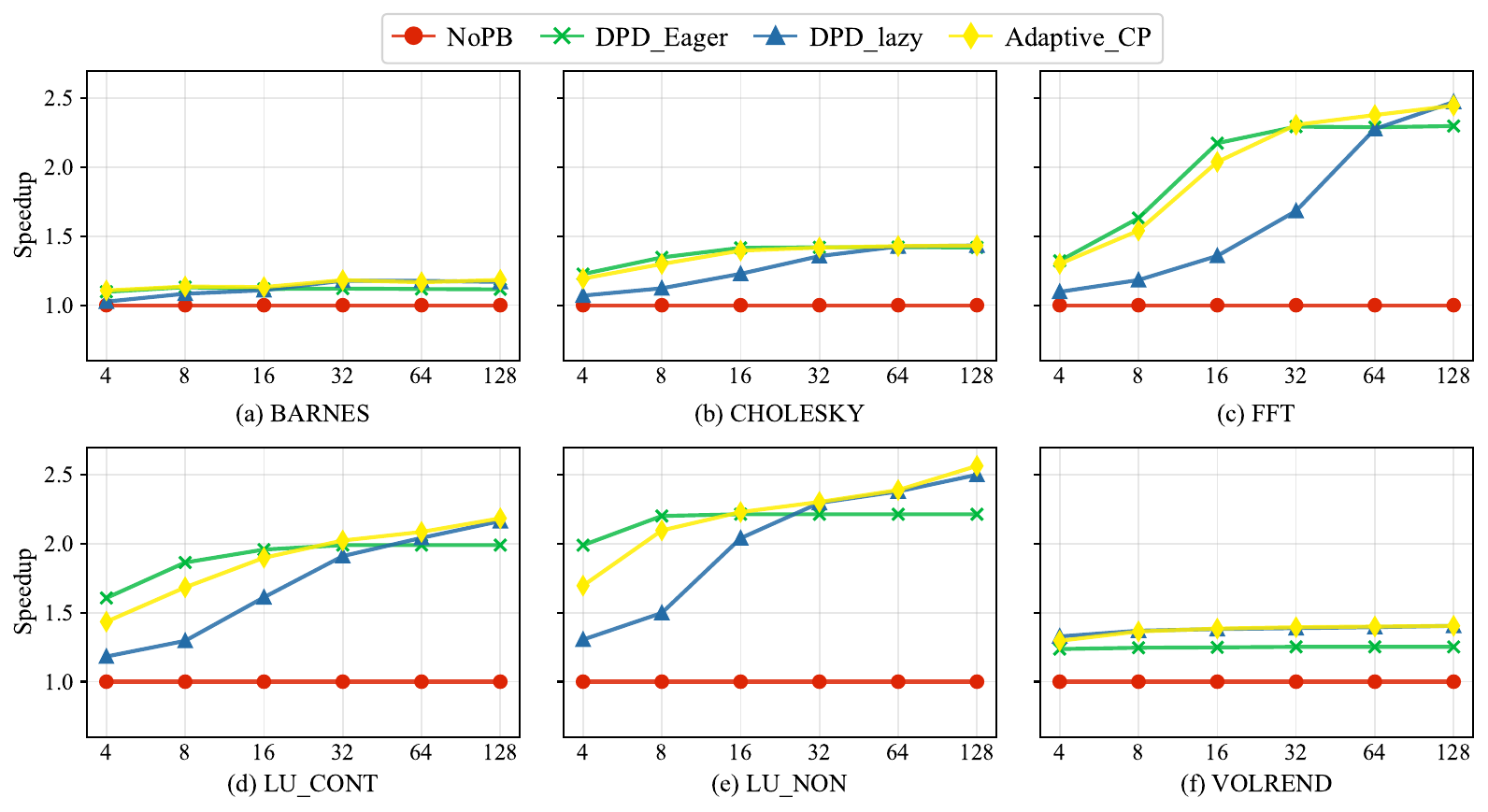}
    \vspace{-2em}
    \caption{Speedup for Splash-4 benchmarks in DPD for different PB entry size at  PCS. X-axis represent number of PB entry exist at PCS.}
    \vspace{-1em}
    \label{fig:pbsize_sensitivity}
\end{figure}

Fig.~\ref{fig:pbsize_sensitivity} shows the speedup of different PCS schemes as the PB size increases. In Fig.~\ref{fig:pbsize_sensitivity}, we exclude RAYTRACE from SPLASH-4 and workloads B, C, and D from YCSB because they show no sensitivity to PB size. Owing to their similar sensitivity trends, we show only BARNES among BARNES, A, and F, and only VOLREND between VOLREND and VOLREND\_NPL in Fig.~\ref{fig:pbsize_sensitivity}. From Fig.~\ref{fig:pbsize_sensitivity}(a) and (f), we observe that BARNES and VOLREND are not very sensitive to PB size, as their performance is not primarily limited by the persist buffer size. As the PB size increases, their speedup improves only slightly. In contrast, FFT, LU\_CONT, and LU\_NON (Fig.~\ref{fig:pbsize_sensitivity}(c)--(e)) are highly sensitive to PB size. When the PB size drops below 16 entries, even DPD\_Eager begins to suffer from the limited number of persist buffer entries. When the PB size is reduced to 4 entries, FFT starts to lose the benefit of having a PB altogether. Increasing the PB size beyond 32 entries does not improve performance for DPD\_Eager; however, for DPD\_Lazy and the adaptive schemes, the speedup continues to increase as the PB size grows. CHOLESKY (Fig.~\ref{fig:pbsize_sensitivity}(b)) is also sensitive to PB size, but less significantly, and it does not benefit from having more than 32 PB entries.


\begin{figure}[!t]
    \centering
    \includegraphics[width=.48\textwidth]{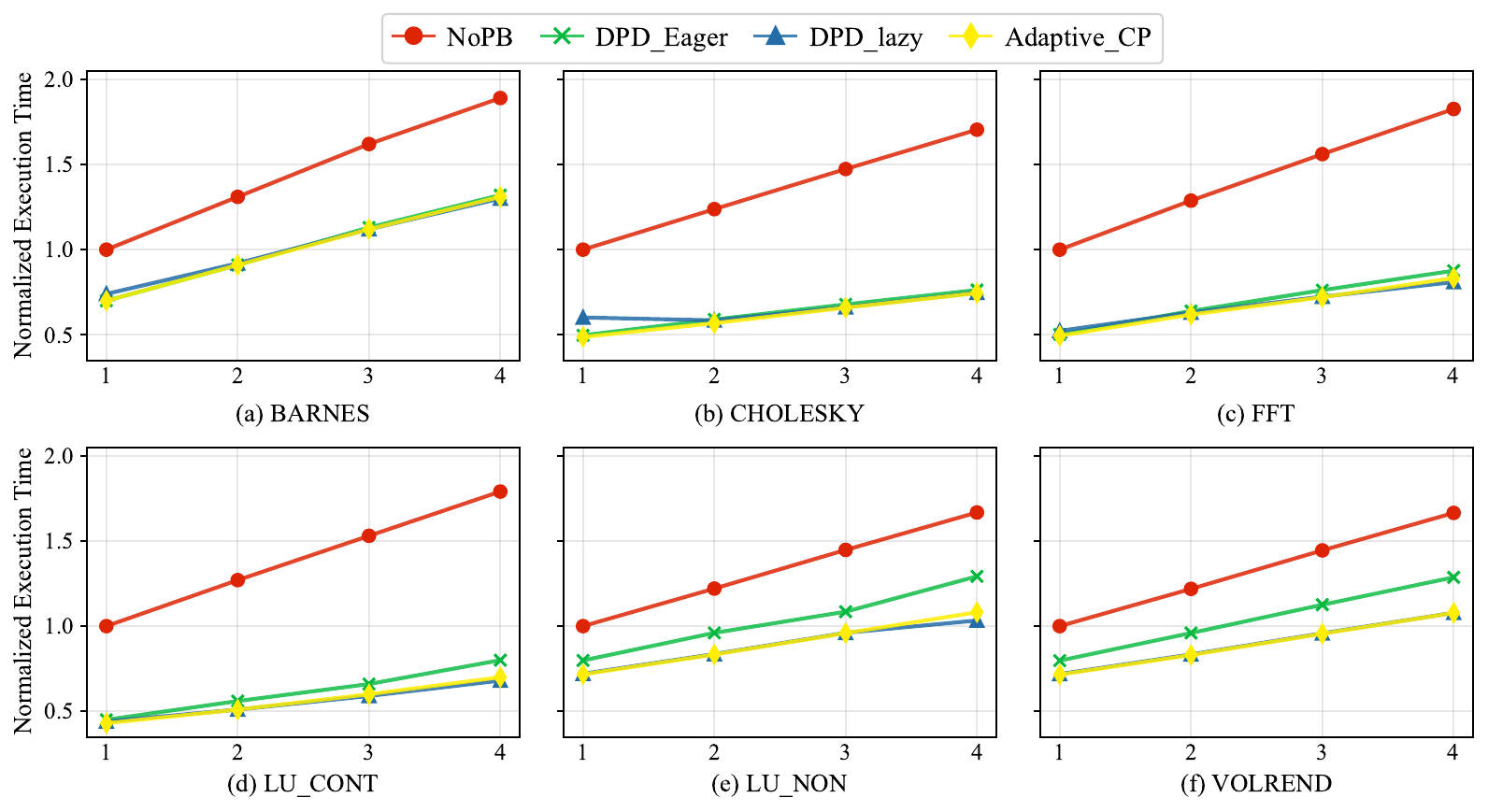}
    \vspace{-2em}
    \caption{Scalability of DPD schemes with increasing number of PCS hop (x-axis) between compute node and memory node. } 
    \vspace{-1em}
    \label{fig:hop_sensitivity}
\end{figure}

\subsubsection{CXL switch hop sensitivity} Fig.~\ref{fig:hop_sensitivity} shows the execution-time sensitivity to an increasing number of CXL switches between the compute node and the memory node. Although PBR routing theoretically supports arbitrary topologies, practical datacenter networks typically use designs with limited maximum hop counts, such as spine--leaf (worst case: 3 hops)~\cite{teng2026hpc_cluster_design,wang2025network_simulator} or dragonfly (worst case: 4 hops)~\cite{shpiner2017dragonfly_plus,garcia2012high_redix_hierarchical_network}. To evaluate how DPD behaves with different hop distances, we test up to four CXL-switch hops for all workloads, assuming that every switch supports persistence. We exclude RAYTRACE and workloads B, C, and D because they show no noticeable improvement with a single PCS hop and similarly gain no benefit as the hop count increases. Owing to their similar execution-time trends as hop count increases, we show only BARNES among BARNES, A, and F, and only VOLREND between VOLREND and VOLREND\_NPL in Fig.~\ref{fig:hop_sensitivity}.


From Fig.~\ref{fig:hop_sensitivity}(b--e), we observe that workloads with significant performance gains from PCS continue to benefit proportionally as more CXL switches are added between the compute and memory nodes.  In DPD\_Lazy, FFT and CHOLESKY suffer from higher write-failure rates, which reduces some of the benefit. Because draining a PBE to the next PCS is faster than draining all the way to memory, the additional hop distance allows PBEs to free up earlier, making the performance difference between DPD\_Lazy and other schemes negligible when more than one PCS hop is present.  Workloads that benefit from keeping data blocks in PCS (Fig.~\ref{fig:hop_sensitivity}(f)) due to high read-hit rates experience increasing performance gaps between DPD\_Eager and the other schemes as hop count grows, because even when the first PCS is a miss, the probability of finding the data in downstream PCS nodes increases. Finally, Fig.~\ref{fig:hop_sensitivity}(a) shows that although BARNES exhibits only modest improvement under DPD\_Lazy with a single PCS hop, the performance gap becomes noticeable as the number of PCS hops increases.




\section{Conclusion}

 In this paper, we propose Distributed Persistence Domain, a disaggregated persistent memory design with switch level data persistency support. We formalize distributed persistence domain and characterize the correctness hazards in distributed persist structures. Building on these insights, we design Persistent CXL Switch, a CXL switch architecture that incorporates persistency support to significantly reduce persist latency, enable read forwarding, and coalesce writes while guarantee correctness and crash consistency across arbitrary CXL topologies and system scales. Our evaluation shows that the proposed design achieves 33\% speedup on average over volatile CXL switch and adaptive read forwarding optimization could increase average speedup up-to 36\%.





\balance
\bibliography{refs}

\end{document}